\documentclass[11pt, oneside,reprint,floatfix,superscriptaddress]{revtex4-1}  	% use "amsart" instead of "article" for AMSLaTeX format
\usepackage{geometry}                		% See geometry.pdf to learn the layout options. There are lots.
\usepackage{graphicx}
\usepackage[justification=raggedright]{caption}
\usepackage[justification=raggedright]{subcaption}
\usepackage{physics}								% TeX will automatically convert eps --> pdf in pdflatex		
\usepackage{amssymb}
\usepackage{float}
\usepackage{color}
\usepackage{multirow}
\begin{document}

%\title{Patterns in Turing Rings of Oscillators:  Effects of Nonlinearity and Multiple Time Scales}
\title{Synchronization Patterns in Geometrically Frustrated Turing Rings}

\author{Daniel Goldstein}
\affiliation{Martin A. Fisher School of Physics, Brandeis University, Waltham, MA}
\author{Michael Giver}
\affiliation{Martin A. Fisher School of Physics, Brandeis University, Waltham, MA}
\affiliation{Dept. of Physics, Iowa State University, Ames, IA}
\author{Bulbul Chakraborty}
\affiliation{Martin A. Fisher School of Physics, Brandeis University, Waltham, MA}

\date{\today}							% Activate to display a given date or no date
\begin{abstract}
Coupled nonlinear oscillators can exhibit a wide variety of patterns. We study the Brusselator as a prototypical autocatalytic reaction-diffusion model. Working in the limit of strong nonlinearity provides a clear timescale separation that leads to a canard explosion in a single Brusselator. In this highly nonlinear regime, it is numerically found that rings of coupled Brusselators do not follow the predictions from Turing analysis. We find that the behavior can be explained using a piecewise linear approximation.
\end{abstract}

\maketitle

\section{Introduction} The emergence of synchronized patterns in systems of coupled oscillators is a phenomenon that controls the function of a wide variety of systems ranging from biological clocks to optoelectronic circuits~\cite{Strogatz}.   Oscillators can synchronize in phase or out of phase depending on the nature of the coupling, and the inherent complexities of the individual oscillators.   A particularly intriguing question is the impact of  frustration on phase and frequency patterns of coupled nonlinear oscillators\cite{PhysRevE.47.864,Levnajic:2012fk}.  Recent experiments in a model of the classic Turing ring of diffusively coupled synthetic ``cells''\cite{Turing} have observed a variety of synchronized patterns with complex phase relationships,  and spatially heterogenous patterns where the oscillation of one or more units was suppressed\cite{PNAS_Nate}.  The experimental model consisted of microfluidically produced surfactant-stabilized emulsions in which ordered rings of aqueous droplets containing the Belousov-Zhabotinsky (BZ) oscillatory chemical reactants are dispersed in oil\cite{PNAS_Nate}.  The dominant phase relation between oscillators was anti-phase ($\pi$ out of phase) even in odd-numbered rings which are geometrically frustrated\cite{PhysRevE.47.864}, and an explosion of patterns occurred at a ring size of five\cite{PNAS_Nate}.
%A classic model of morphogeneis is the Turing  ring of  diffusively-coupled discrete cells\cite{Turing}.   In this model, each cell is characterized by a group of chemical reactions that influence each other through the diffusion of selective species controlled by the cell membrane.   In Turing's original paper, linear stability analysis was used to predict a host of patterns.   Very recently, these predictions have been experimentally verified in a model system of  synthetic ÒcellsÓ: microfluidically produced surfactant-stabilized emulsions in which aqueous droplets containing the BelousovÐZhabotinsky (BZ) oscillatory chemical reactants are dispersed in oil\cite{PNAS_Nate}.   These experiments reported {\color{red}  oscillating patterns with complex phase relationships between the droplets, and stationary states.  Some of these patterns were not predicted by the linear stability analysis and  were ascribed to the presence of heterogenities. }

{Our work is motivated by these experiments  and the broader question of  the emergent behavior of a set of coupled, strongly nonlinear oscillators  in a geometrically frustrated array.   In such an environment,  coupled oscillators can relieve the frustration by synchronizing in phase, suffering oscillator death\cite{OD}, or exhibiting complex spatio-temporal patterns that accommodate the frustration but preserve the anti-phase relations.  We find that in the regime of strong nonlinearities, the phase space is dominated by complex spatio-temporal patterns whereas weakly nonlinear oscillators synchronize in phase or suffer oscillator death.  

The systems we study are Turing rings where the chemistry of the ``cells''  is modeled by one of the simplest autocatalytic reactions: the Brusselator\cite{Brusselator}.  These Brusselators are coupled through the diffusion of the inhibitory species, which mimics the diffusion in a number of situations including the microfluidic emulsion\cite{PNAS_Nate}.    Coupled Brusselators have been studied extensively\cite{CB1}\cite{Osipov}\cite{Ilya_book}, however, attention has been mainly focused on the regime where the nonlinearity is weak\cite{Osipov} or moderate.    In the weakly nonlinear regime,  there is a large separation of time scales with the activator being the fast species\cite{Osipov} and  a canard explosion leads to relaxation oscillators.  We focus on the regime of {\it strong nonlinearity},  which is also characterized by a large separation of time scales and a canard explosion\cite{Baer_singular} leading to relaxation oscillations with a qualitatively different form of the limit cycle~\cite{Strogatz}. In this regime, the {\it inhibitor} is the fast species, and the slow variable is the total concentration of chemicals: activator plus inhibitor.   We show that coupling Brusselators in this strongly non-linear regime leads to a wide variety of spatio-temporal patterns that are characterized by a strong preference for neighboring oscillators to be $\pi$ out of phase.    

We  characterize the attractor space of rings of oscillators, coupled via the inhibitory species,  through numerical analysis and application of  a piecewise linear approximaiton (PLA)  that is valid in the large nonlinearity regime.   Analysis of the model based on PLA shows that in this highly nonlinear regime  the dynamics of the coupled Brusselators can be represented as oscillators traveling on branches of the limit cycle  separated by near-instantaneous jumps between these branches. In one branch of the limit cycle, the oscillators are repulsively coupled while on the other they are attractively coupled.  The sequence of jumps between the branches characterizes  the patterns observed numerically.  This representation, in terms of jumps provides an understanding the rich attractor space in the highly-nonlinear regime, and we suggest that this technique will be useful for other reaction-diffusion systems with strong nonlinearities\cite{Piecewise}.

This paper is organized as follows. First,  we present an analytical investigation of a single  Brusselator  with emphasis on the strongly nonlinear regime.  We perform a change of variables that clearly demonstrates the  separation of time scales, and present a review of the standard Turing analysis on a ring of $N_r$ Brusselators, using this representation. This is followed by a numerical investigation, which  examines the phase and frequency synchronization properties of rings with $N_r$  ranging  from 2 to 5. Finally we apply the PLA to the Brusselator in the regime of large nonlinearity in order to understand the origin of the frustration in $N_r$-rings, and the patterns that emerge in numerical solutions. 

\section{The Single Brusselator}
In this section, we review the properties of the single Brusselator in order to clearly identify the regions of weak and strong nonlinearities and the time-scale separation that drives the physics in the regime of large nonlinearity.

{ The Brusselator\cite{Galla} is a model of an auto catalytic chemical reaction that was first proposed by Ilya Prigogine at the Free University of Brussels in the 1960s. It is a two species model, with an activator $X$ and an inhibitor $Y$. There are several ways to write the set of reactions~\cite{OGbruss, brussref1, brussref2, Galla}.  We choose the representation  below~\cite{Galla, OGbruss} since this form is particularly amenable to a systematic expansion about the mean field limit\cite{Galla}, and we would like to explore these effects in the near future.
\begin{align} \label{eq:rxn}
A&\rightarrow X+A&N\nonumber\\
X&\rightarrow0&n_{x}\nonumber\\
X+B&\rightarrow Y+B&bn_{x}\\
2X+Y&\rightarrow3X&cN^{-2}n_{x}^2n_{y}\nonumber
\end{align}
In this representation, the populations of the $A$ and $B$ species are constant in time, and their only role is to set the reaction rates\cite{Galla}.   The first reaction describes the creation of  $X$ molecules with a rate  proportional to $N$, the number of $A$ molecules.  The first two reactions, in isolation, ensure that the mean number of $X$ molecules over time is given by $N$, which  acts as a measure of the system size.   The other parameters appearing in the reactions  are as follows:  $b$ is the rate of exchange of activator to inhibitor, $c$ controls the  rate of the nonlinear reaction in which the inhibitor transforms back into the activator, and $n_x$ and $n_y$ are the number of $X$ and $Y$ molecules, respectively.   }

A well-mixed system,  in the limit of large system size ($N$),  is described by well-known rate equations in which the rates of change of  the concentrations of $X$ ($x$) and $Y$ ($y$) are obtained by using the law of mass action.  The rate equations can also be obtained systematically through a Van-Kampen expansion\cite{Galla}, and that approach demonstrates the mean field nature of the equations:
\begin{align}\label{eq:rate}
\dot{x}&=1-x(1+b-cxy)\nonumber \\
\dot{y}&=x(b-cxy)
\end{align}
where $x=X/N$ and $y=Y/N$
In this paper, we focus exclusively on the mean field equations. The presence of a rich attractor space and metastability suggests that intrinsic fluctuations, neglected in the mean field limit, can modify patterns qualitatively\cite{Giver_Danny_me,other_dg_mg_bc}.  

The above set of nonlinear first order differential equations (Eq.~\ref{eq:rate}) has a unique fixed point at $(\bar{x},\bar{y})=(1,(\delta+c+1)/c)$. Where $\delta \equiv b-c-1$ is the control parameter for the Hopf instability in the system. By examining the behavior of these equations near this fixed point one can show that the parameter space of $\delta$ and $c$ is broken up into the four regions as shown in Figure \ref{fig:param}. In regions (1) and (2) where $\delta<0$, the fixed point is stable:  in regions (3) and (4),  the fixed point is unstable. The eigenvalues of the Jacobian matrix are complex in regions (2) and (3) indicating the fixed point lies at the end of a stable and unstable spiral, respectively. In region (3) a stable limit cycle emerges from the fixed point with an amplitude that grows as $\sqrt{\delta}$  for small  $\delta$.
\begin{figure}[H]
   \includegraphics[width=\linewidth]{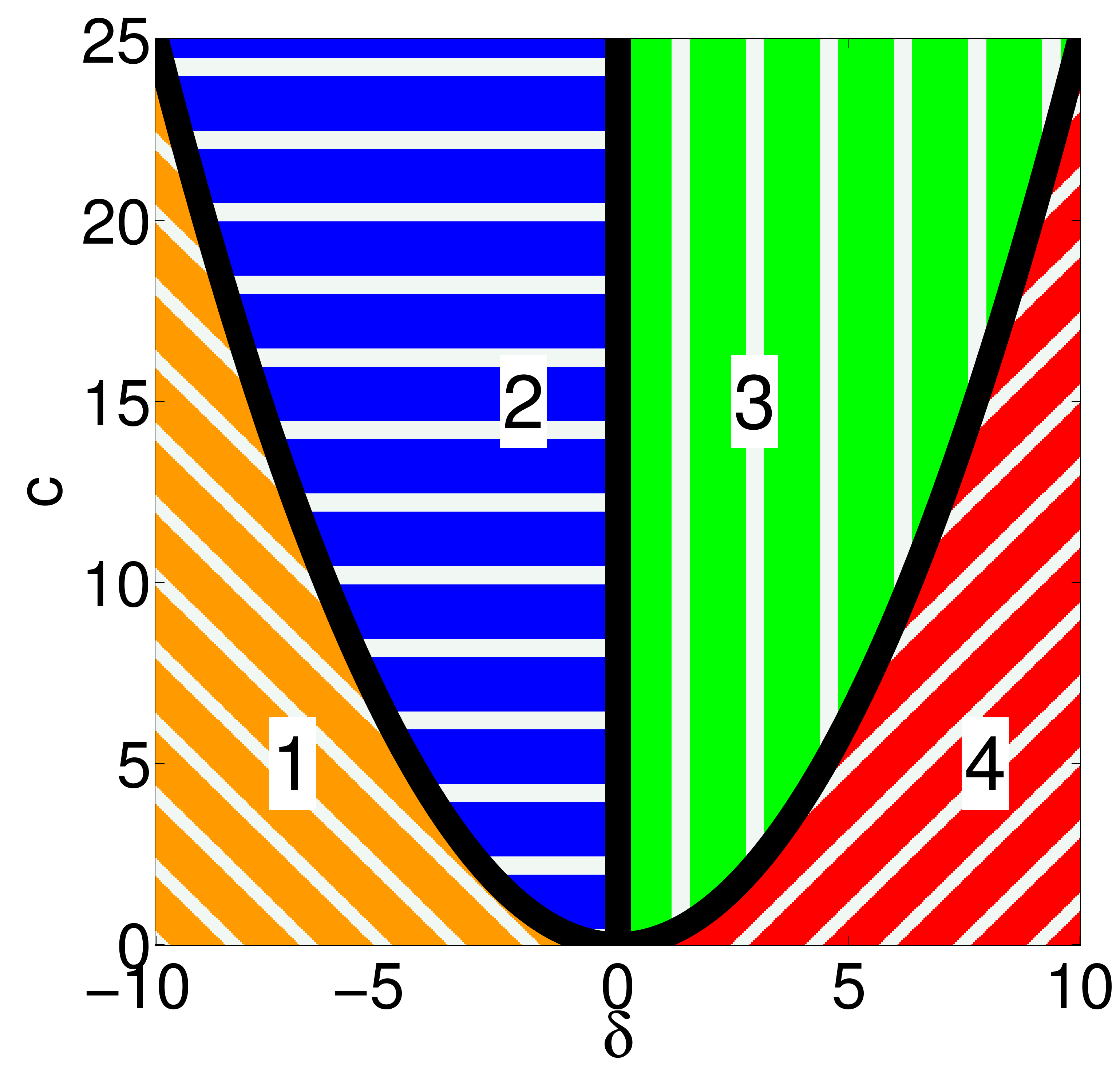}
   \caption{(Color Online) Regions of stability about the fixed point in b-c parameter space. In regions (1) and (2) the fixed point is stable while in (3) and (4) it is unstable. Additionally, regions (2) and (3) have oscillatory behavior around the fixed point.}
   \label{fig:param}
\end{figure}

As $c$ is increased beyond some threshold value, however, the dynamics of the Brusselator changes qualitatively: it enters a regime  characterized by  a large separation of time scales. This difference in time scales becomes apparent under the change of variable $u=y$ and $v=x+y$.  In this representation,  the rate equations become.
\begin{align}\label{eq:sing}
\dot{u}&=(v-u)([1+\delta]+c[1-u(v-u)])\nonumber \\
\dot{v}&=1-v+u
\end{align}
It can be easily seen  from these equations that for  $c > > 1$, $u$ is the fast variable and $v$ is the slow variable: $x+y \simeq \rm {constant}$ when $\dot{u}\not=0$. In this regime,  the Hopf bifurcation acquires a singular character\cite{Baer_singular}, and at $\delta \gtrsim 1/c$, the system passes though a canard point where the amplitude of the limit cycle jumps sharply. Figures ~\ref{fig:smallcphase}-\ref{fig:largecphase} show the shape of the limit cycle for a fixed $\delta$ as a function of $c$. Fig.~\ref{fig:largecphase} compares amplitudes for two different close values of $\delta$ at a large c. The dependence of the canard explosion on the parameter $c$ can be seen in Figures~\ref{fig:amps}.  It is important to note here that in mean field the single Brusselator is a monostable system with its limiting behavior either being a fixed point, the small limit cycle or the large limit cycle. 
%,  and $\delta << c$
%In the next section, we present numerical results for oscillating patterns of coupled oscillators, and in section 3, we present analysis of a piecewise linear approximation (PLA) to the Brusselator at high $c$, and show that it captures most features of the full numerical simulations.  The PLA provides and elegant framework for understanding the multi-attractor space of coupled, highly nonlinear Brusselators.

% As we show below when the system is simulated with intrinsic noise switching and metastability occurs. % This leads to a psudo-conservation law that $v$ is constant when the system in not on the nullcline of $u$. Of note is that this time scale separation disappears when $b~c~1$, and $v$ is the fast variable for $b,c<<1$.  

\begin{figure*} 

%        \quad
%                \begin{subfigure}[t]{0.3\linewidth}
%                \includegraphics[width=\linewidth]{TimeSepHugeC.png}
%                \subcaption{Red: Phase portrait for c=100 b=101.2 in the u - v plane this is below the canard point. Teal: Phase portrait for c=10 b=102 in the u - v plane this is above the canard point. The nullclines and velocity arrows are calculated below the canard point but do not qualitatively change about the canard point. The scale has changed to accommodate the increased amplitude. The nullclines have not qualitatively changed from the lower values of c.}
%               % \label{fig:3 brusshigh}
%        \end{subfigure}%
        %
        \begin{subfigure}[t]{0.45\linewidth}
        \includegraphics[width=\linewidth]{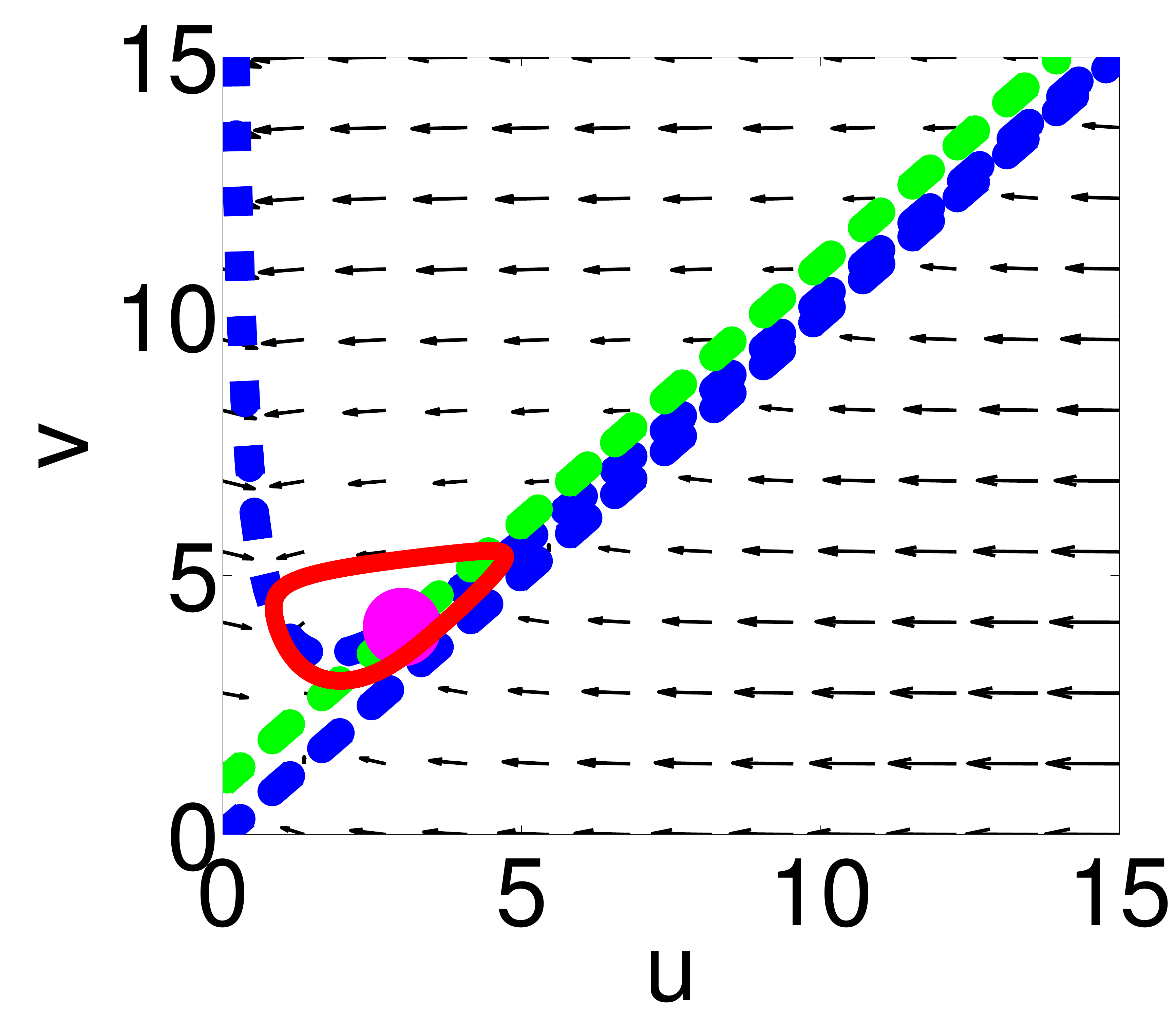}
        %\centering
        \caption{(Color) Nullclines (dashed blue - $u$, dashed green -$v$), velocity field (black) and limit cycle (red) for $c=1$, $\delta$=1. There are two branches of the the $u$ nullcline that asymptotically approach each other. The limit cycle grows out of a Hopf bifurcation of the fixed point (magenta dot).}
        \label{fig:smallcphase}
        \end{subfigure}
        \quad
        \begin{subfigure}[t]{0.45\linewidth}
        \includegraphics[width=\linewidth]{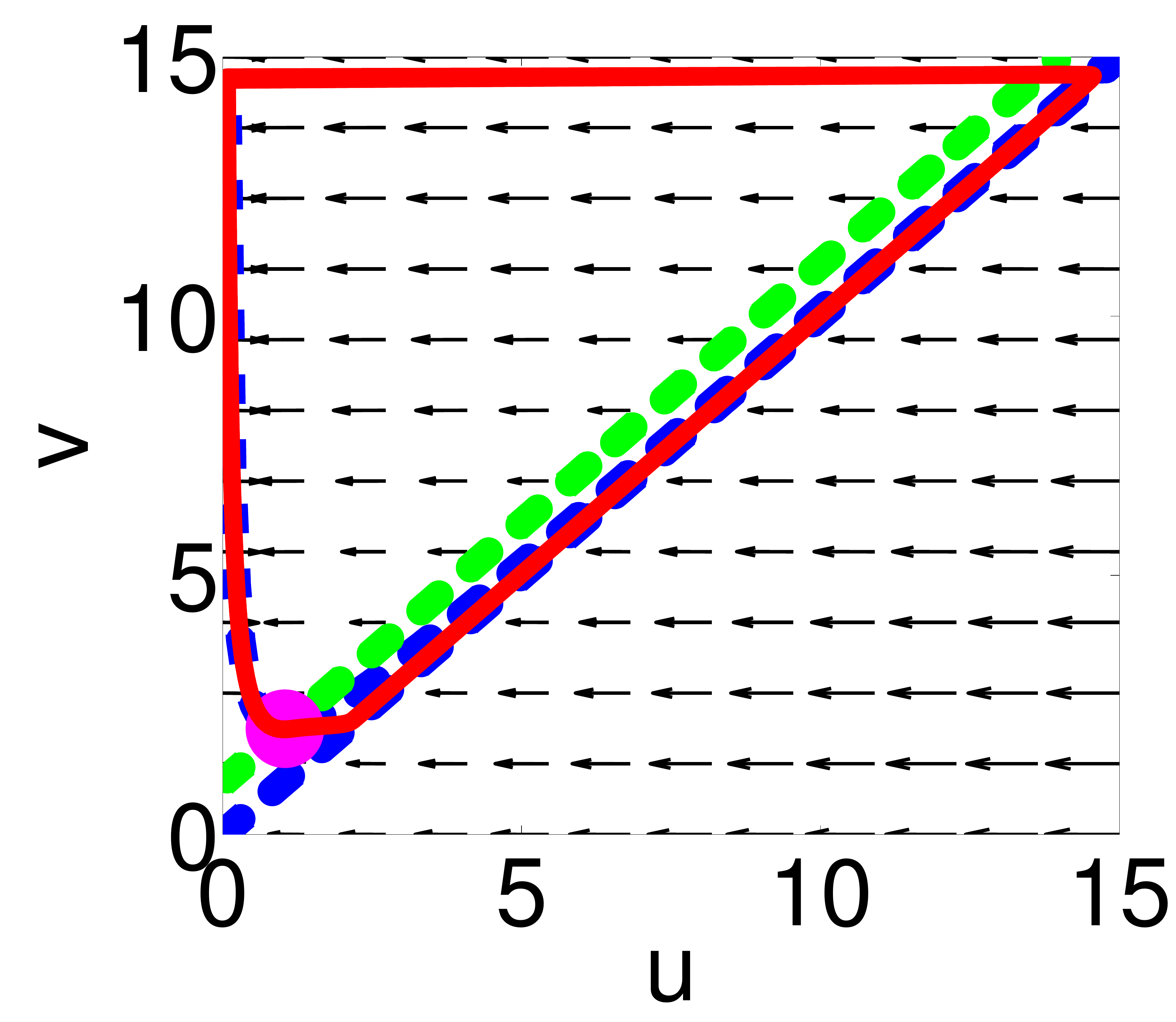}
                \caption{(Color) Nullclines (dashed blue - $u$, dashed green -$v$), velocity field (black) and limit cycle (red) for $c=50$, $\delta$=1. The asymptotic approach of the two branches of the the $u$ nullcline is clearer at this value of $c$, and the limit cycle grows out of a {\it singular} Hopf bifurcation of the fixed point (magenta dot). For this value of $c$, the   limit cycle follows the nullclines of $u$ very closely.}
                \label{fig:largecphase}
        \end{subfigure}

                \begin{subfigure}[t]{0.45\linewidth}
                \includegraphics[width=\linewidth]{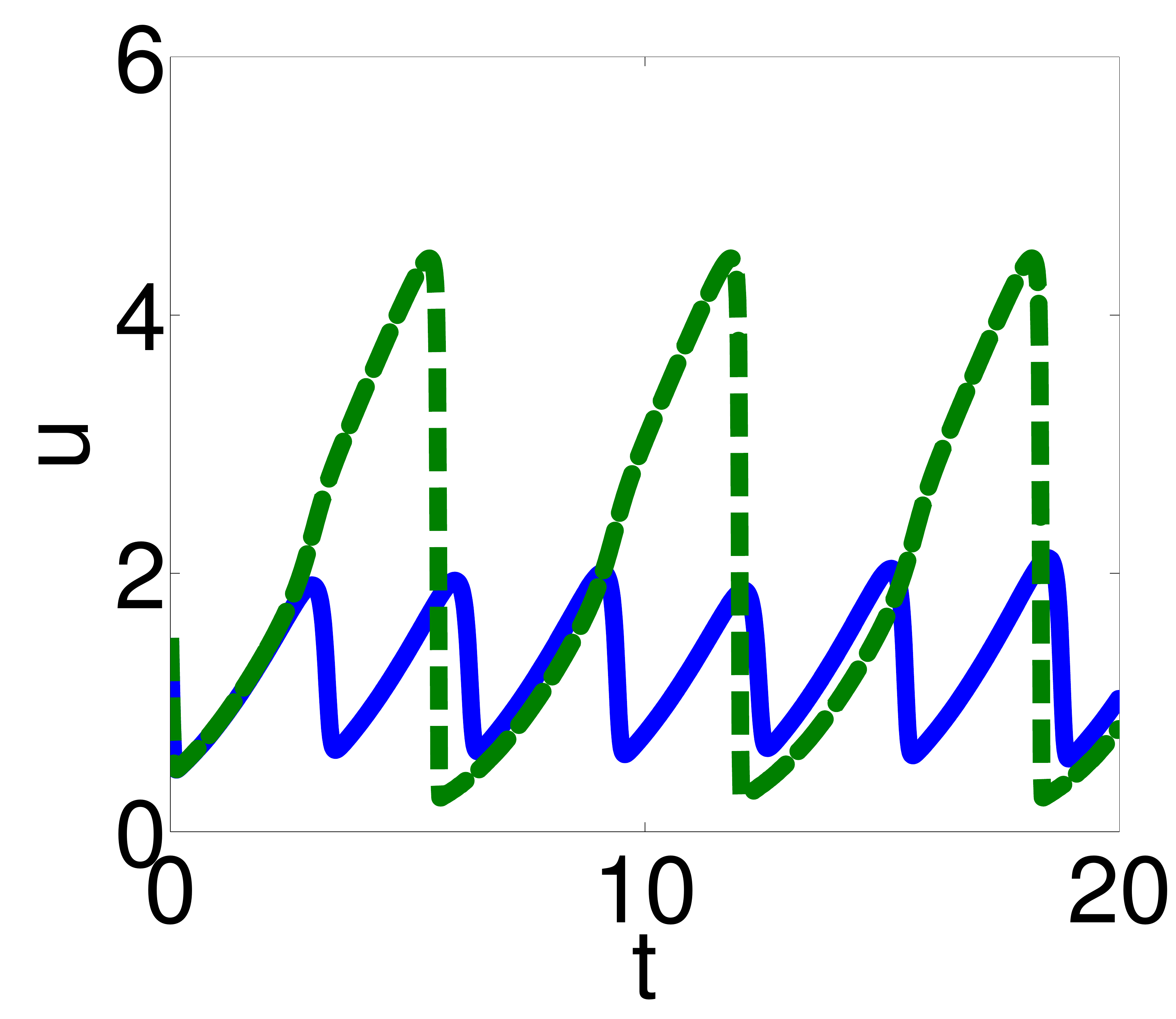}
                \caption{(Color Online)Blue (solid): Time series of u for c=10 $\delta$=0.3:  this is below the canard point. The limit cycle has a period of about 2.5 and an amplitude of about 0.7. Green (dashed): Time series of u for c=10 $\delta$=0.4:  this is above the canard point. The limit cycle has a period of about 5.2 and an amplitude of about 4.}
                \label{fig:canardtime}
        \end{subfigure}
        \quad
                                \begin{subfigure}[t]{0.45\linewidth}
                \includegraphics[width=\linewidth]{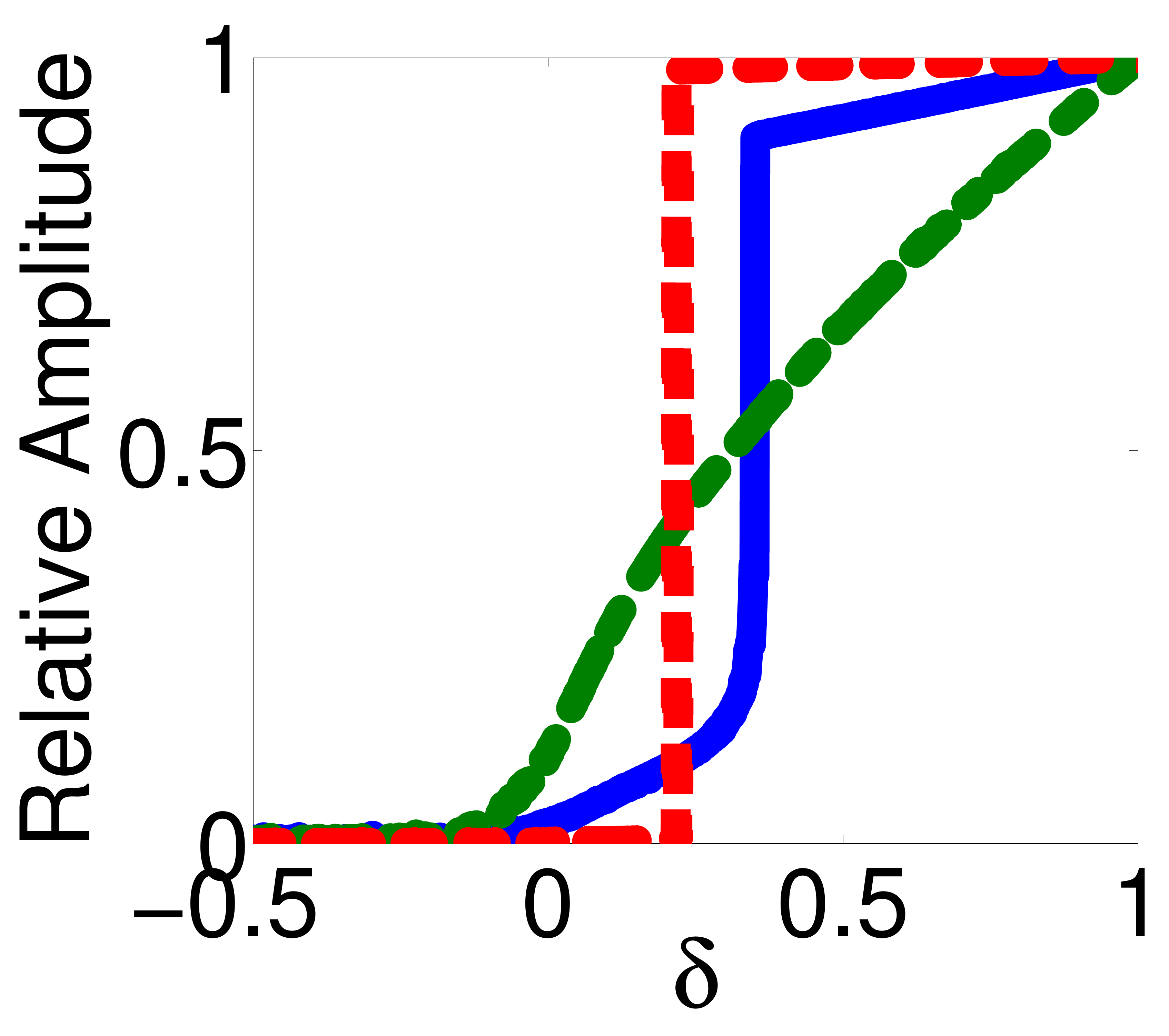}
                \caption{(Color Online)Green (dash - dot), Blue (solid) and Red (dotted) are the scaled amplitudes of the limit cycle for c=1, 10 and 100 respectively. These scaled amplitudes are plotted against $\delta=b-c-1$, and show that $c=1$ has no canard explosion  in this range of $\delta$ while $c=10$ and $100$ do. } 
                \label{fig:amps}

        \end{subfigure}
        
        %
%        \quad
%                \begin{subfigure}[t]{0.3\linewidth}
%                \includegraphics[width=\linewidth]{TimeSepSeriesLargeC.png}
%                \caption{Blue: Time series of u for c=100 b=101.2 this is below the canard point. The limit cycle has a period of about 0.8 and an amplitude of about 0.6. Green: Time series of u for c=100 b=102 this is above the canard point. The limit cycle has a period of about 28 and an amplitude of about 27.}
%                %\label{fig:6 brusshigh}
%        \end{subfigure}%

%\begin{figure*}[H]
%   \centering
%   \includegraphics[width=.5\linewidth]{n=1low.png}
%   \caption{Limit cycle for a single brusselator when c=1 b=2.2.}
%   \label{fig:lowclim}
%\end{figure*}

                        \caption{}
        \label{fig:brusshigh}
        \end{figure*}

\section{Ring of  Coupled  Brusselators}
The previous section provided an overview of the properties of the single Brusselator in the regime of high non linearity.  In this section, we extend our analysis to a ring of Brusselators coupled through $Y$, the inhibitor species. To investigate this system, we introduce a "hopping"  term for the inhibitor to the list of reactions, which translates to a diffusive coupling term in the rate equations. The list of reactions now becomes:
\begin{align}
A&\rightarrow X_i+A&N\nonumber\\
X_i&\rightarrow0&n_{x_i}\nonumber\\
X_i+B&\rightarrow Y_i+B&bn_{x_i}\\
2X_i+Y_i&\rightarrow3X_i&cN^{-2}n_{x_i}^2n_{y_i}\nonumber\\
Y_i&\rightarrow Y_{i+1}&dn_{y_i}\nonumber\\
Y_i&\rightarrow Y_{i-1}&dn_{y_i}&,\nonumber
\end{align}
where $i$ labels the individual Brusselators. The corresponding rate equations are: 
\begin{eqnarray}\label{eq:ratec}
\dot{x_i}&= &1-x_i(1+b-cx_i y_i) \\ 
\dot{y_i}&= & x_i(b-cx_iy_i)+d(y_{i+1}-2 y_{i} + y_{i-1}) \nonumber 
\end{eqnarray}
Where $d$ is the parameter that controls the coupling strength. In the $u$ and $v$ variables, the equations are:  
\begin{align}\label{eq:singc}
\dot{u_i}&=(v_i-u_i)(1+\delta)\\
	     &\;\;+c[(v_i-u_i)-u_i(v_i-u_i)^2]\nonumber\\
	     &\;\;+d(u_{i+1}-2 u_{i} + u_{i-1})\nonumber \\
\dot{v_i}&=1-v_i+u_i+d(u_{i+1}-2 u_{i} + u_{i-1}) \nonumber
\end{align}
{ The original diffusive coupling is realized in this representation as a coupling of the fast variables, $u$, and an additional driving term for the slow variable, $v$}.
\subsection{Linear stability analysis of Ring}
%{\bf Discuss Turing analysis and point out the $q=0$ and $q=\pi$ mode competition.   Then put in the table based on considering only the fastest growing instability}
In order to understand the emergence of patterns  in a  ring of $N_r$ Brusselators we perform a standard linear stability analysis of the homogeneous fixed point; that is the fixed point of a Brusselator without coupling. The analysis is summarized here for completeness.  We  linearize each Brusselator about the fixed point at $(\bar{u}_i,\bar{v}_i)=(\frac{1+\delta+c}{c},1+\frac{1+\delta+c}{c})$.  For the uncoupled system, the deviation from the fixed point obeys the equation: 
\begin{align}
\mqty[\dot{u}\\\dot{v}]=J\mqty[u\\v]
\end{align}
Where J is the jacobian of this system evaluated at the fixed point:  $J=\mqty(1+\delta&-(1+\delta+c)\\1&-1)$.
Adding the diffusive coupling leads to:  
\begin{align}
\mqty[\dot{\mathbf{u}}\\\dot{\mathbf{v}}]=(J+D)\mqty[\mathbf{u}\\\mathbf{v}]
\label{eq:vectorturing}
\end{align}
where D is the coupling matrix for $\mathbf{u}$ and $\mathbf{v}$
We now look for solutions to the linear equation with  the form: 
\begin{align}
\mathbf{u}=\mathbf{u}_q e^{\sigma_q t }e^{iqs}
\end{align}
Where now $\sigma_q$ is the growth rate, q is the wavenumber and $s=1,...,N_r$ is the oscillator number.  Substituting this form into Eq.~\ref{eq:vectorturing}, 
\begin{align}
&\sigma_q\mqty(\mathbf{u}\\
\mathbf{v})=J\mqty(\mathbf{u}\\\mathbf{v})\\
&+d\mqty(e^{iq(s-1)}-2e^{iqs}+e^{iq(s+1)}&0\\e^{iq(s-1)}-2e^{iqs}+e^{iq(s+1)}&0)\mqty(\mathbf{u}\\\mathbf{v})\nonumber
\end{align}

The allowed values of  $q$ are $q=\frac{n2\pi}{N_r}$ with $n$ an integer, and $\sigma_q$ is determined by:
\begin{align}
\sigma_n\mqty(\mathbf{u}\\\mathbf{v})=\hat{J_n}\mqty(\mathbf{u}\\\mathbf{v}) ~,
\end{align}
with  $\hat{J_n}$ given by:
$$\hat{J_n}=\mqty(1+\delta-4d\sin^2(\frac{n\pi}{N_r})&-(1+\delta+c)\\1-4d\sin^2(\frac{n\pi}{N_r})&-1) ~ .$$
Stability of the fixed point  requires that the real part of both eigenvalues is negative which is satisfied when $\Tr(\hat{J})$ is negative and $\det(\hat{J})$ is positive for all $n$. Thus the requirements for stability are:
\begin{align}
\delta-4d\sin^2(\frac{n\pi}{N_r})&<0\\
c-4d(c+\delta)\sin^2(\frac{n\pi}{N_r})&>0
\end{align}
{ We are interested in cases where the coupling $d<<1$, and in this regime, the fixed points are stable for $\delta<0$.} For $\delta>0$ the fastest growing mode is the one for which  $\sin^2(\frac{n\pi}{N_r})=0$, therefore,  the $q=0$ mode. This suggests that the Brusselators in a ring will always oscillate in phase with one another.   In the following sections, we present results of numerical simulations for rings with $N_r =2 ~- ~ 5$.  We have carefully investigated the phase-diagram  in $d-\delta$  space for $N_r=2,5$. For the $N_r=2$ system we focus on explaining the patterns using the piecewise linear approximation. For $N_r=5$ we emphasize the complexity of the patterns that form and the regions that they form in.} 

%Thus the only predicted Turing modes are the ones with extremely long wavelength, stationary (Turing (a)) or all Brusselators oscillating in phase (Turing (b)) \cite{Turing}

% Requires the booktabs if the memoir class is not being used
%\begin{table*}[htbp]
%   \centering
%   %\topcaption{Table captions are better up top} % requires the topcapt package
%   \begin{tabular}{|p{.1\linewidth}|p{.25\linewidth}|p{.25\linewidth}|p{.25\linewidth}|} % Column formatting, @{} suppresses leading/trailing space
%   \hline
%  Turing Pattern & Description & Expected in our system & Present in our system\\ \hline
%   a & Stationary waves of extremely long wavelength & $\delta<0$ & For all parameters below the fixed point and with small coupling\\ \hline
%   b & Oscillatory waves of extremely long wavelength & $\delta>0$ & For small y-coupling below the canard and small enough coupling compaired to $\delta$ \\ \hline
%   c & Stationary waves of extremely short wavelength & & For high y-coupling\\ \hline
%   d & Stationary waves of finite wavelength & & $\times$\\\hline
%   e & Oscillatory case with a finite wavelength & Needs a 3 variable model & $\times$\\\hline
%   f & Oscillatory case with extreme short wavelength & Needs a 3 variable model & For small y-coupling above the canard \\ \hline
%   \end{tabular}
%   \caption{}
%   \label{tab:booktabs}
%\end{table*}

%Going through the Turing analysis for this system we find the the only precidcted mode is mode c, that is to say stationary waves of extreemly short wavelength

\section{Numerical Results}
\subsection{Two Oscillators} In this section, we  present a numerical study of the synchronization of two Brusselators, focusing primarily  on the large $c$ regime.   From the linear stability analysis, the oscillators  are always expected to synchronize in phase.  Simulating this system in the mean field limit we find that when $c=1$ that two Brusselators will synchronize in phase with one another  as the Turing analysis predicts. In the $c\gg1$ regime,  we find that when the system is above the canard point - in the large limit cycle - the Brusselators oscillate out of phase with one another, while below the canard explosion they  oscillate in phase with one another. %The trajectories from the Gillespie algorithm confirm these results, giving noisy out of phase large limit cycles and noisy in phase small limit cycles.

For $N_r=2$, we observe  four different types of behavior - homogeneous stable fixed points, inhomogeneous stable fixed points (oscillator death), in phase oscillations and out of phase oscillations. We notice that if the coupling strength $d$ is above a critical value $d_c$  then the Brusselators experience oscillator death~\cite{OD}: another, nontrivial pair of fixed points becomes real and stable through a pitchfork bifurcation for $d>d_c \equiv \frac{4c}{(c+\delta+2)^2}$\label{sec:oscdeath}. This type of behavior, however, is not the focus of this paper. There are also regions where the behavior is dependent on initial conditions, these are enumerated in Table \ref{tab:phases}. It is in these regions that hysteresis can be observed.  A complete portrait of the attractor space in $d-\delta$ space will be discussed in the context of the PLA, and is shown in Fig.~\ref{fig:phasecomp}. 

\begin{table}[htp]
    \begin{tabular}{| l | l |}
    \hline
    Region & Possible states\\\hline
    1 & Homogeneous fixed point\\\hline
    \multirow{2}{*}{2} 	& Homogeneous fixed point,\\
    				& Out of phase oscillations.\\\hline
    \multirow{2}{*}{3}  	& Homogeneous fixed point,\\
    				& Oscillator death.\\\hline
    4 & Out of phase oscillations.\\\hline
    5 & Oscillator death.\\\hline
    \multirow{2}{*}{6} 	& Out of phase oscillations,\\
    				& In phase oscillations.\\\hline
    \multirow{2}{*}{7} 	& Oscillator death.\\
    				& In phase oscillations.\\\hline
    8 & In phase oscillations.\\\hline
    \end{tabular} 
    \caption{States observed in a numerical investigation of the 2 - ring. A complete phase diagram with numbered regions is found in Fig.~\ref{fig:phasecomp}}
   \label{tab:phases}
\end{table}
%
%\begin{figure*}[ht]
%   \begin{subfigure}[t]{.45\linewidth}
%   \includegraphics[width=\linewidth]{2phaseplot}
%   \caption{Phase {\color{red} diagram}  for 2 coupled Brusselators for $c=50$. Regions and possible behaviors are enumerated to the right. If multiple behaviors exist for a single regime of parameters then they are accessible through choice of initial condition. }
%   \label{fig:twophase}
%   \end{subfigure}
%    \begin{subfigure}[T]{.45\linewidth}
%    \begin{tabular}{| l | l |}
%    \hline
%    Region & Possible states\\\hline
%    1 & Homogeneous fixed point\\\hline
%    \multirow{2}{*}{2} 	& Homogeneous fixed point,\\
%    				& Out of phase oscillations.\\\hline
%    \multirow{2}{*}{3}  	& Homogeneous fixed point,\\
%    				& {\color{red}Oscillator death}.\\\hline
%    4 & Out of phase oscillations.\\\hline
%    5 & {\color{red}Oscillator death}.\\\hline
%    \multirow{2}{*}{6} 	& Out of phase oscillations,\\
%    				& In phase oscillations.\\\hline
%    \multirow{2}{*}{7} 	& {\color{red}Oscillator death}.\\
%    				& In phase oscillations.\\\hline
%    8 & In phase oscillations.\\\hline
%    \end{tabular}
%    
%    \end{subfigure}
%    \label{fig:2phase}
%\end{figure*}
Summarizing, when the system is in the small limit cycle,  in-phase oscillations occur, as predicted by the linear stability analysis. When the system is in the large limit cycle,  the linear stability analysis fails for some regions of parameter space, and out of phase oscillations can be observed. These out of phase oscillations are expected to lead to complex patterns when  the oscillators are arranged in a geometry that is frustrated:   the oscillators cannot be $\pi$ out of phase with all of their nearest neighbors.  From here on we only look at Brusselators with large nonlinearity:  specifically,  $c=50$ and below the coupling at which they stop oscillating ($d<d_c$).

\subsection{Three Brusselators}
\begin{figure}[htp]
        \begin{subfigure}[b]{0.45\linewidth}
\includegraphics[width=\linewidth]{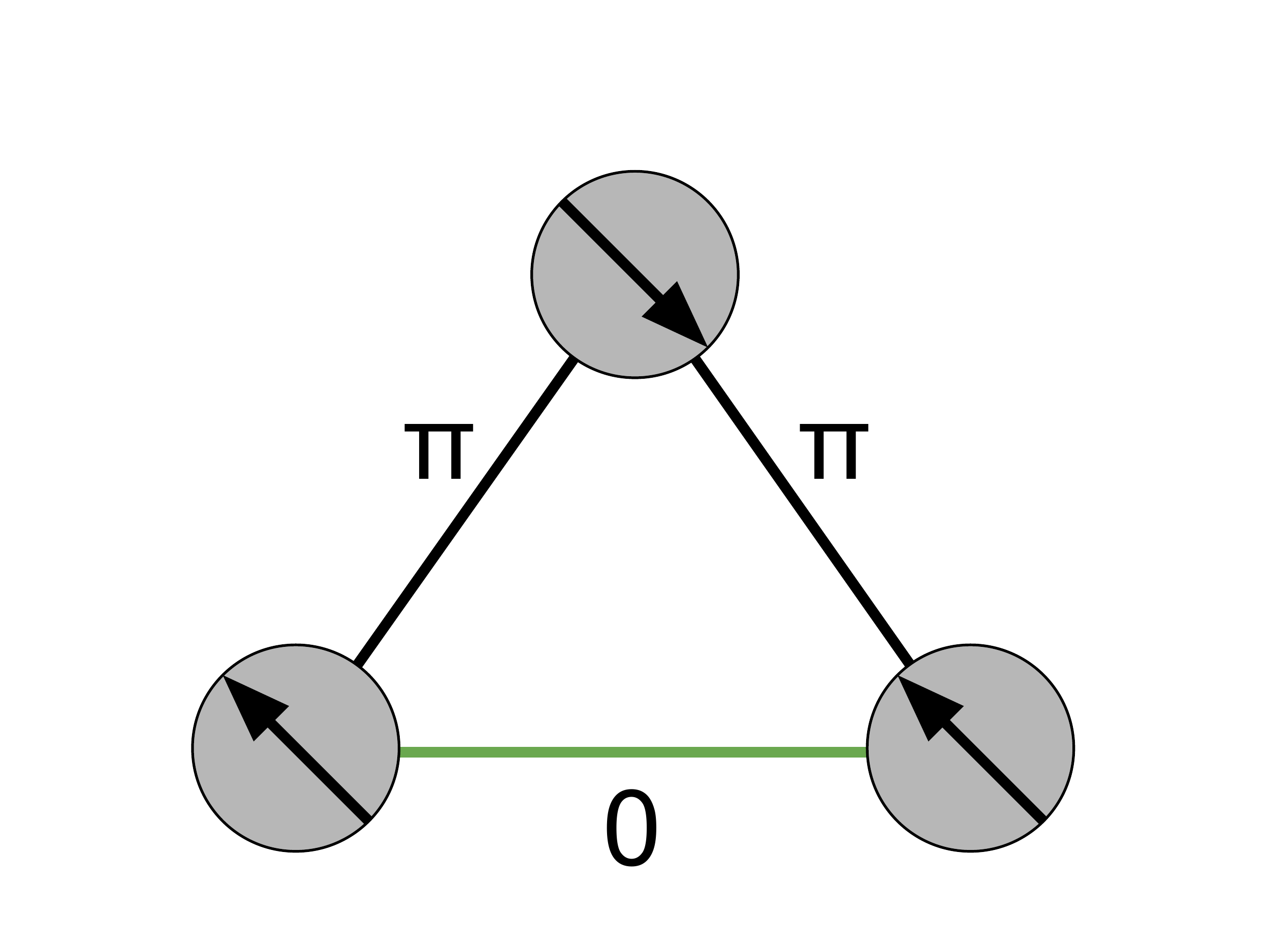}
        \end{subfigure}%
                \begin{subfigure}[b]{0.45\linewidth}
\includegraphics[width=\linewidth]{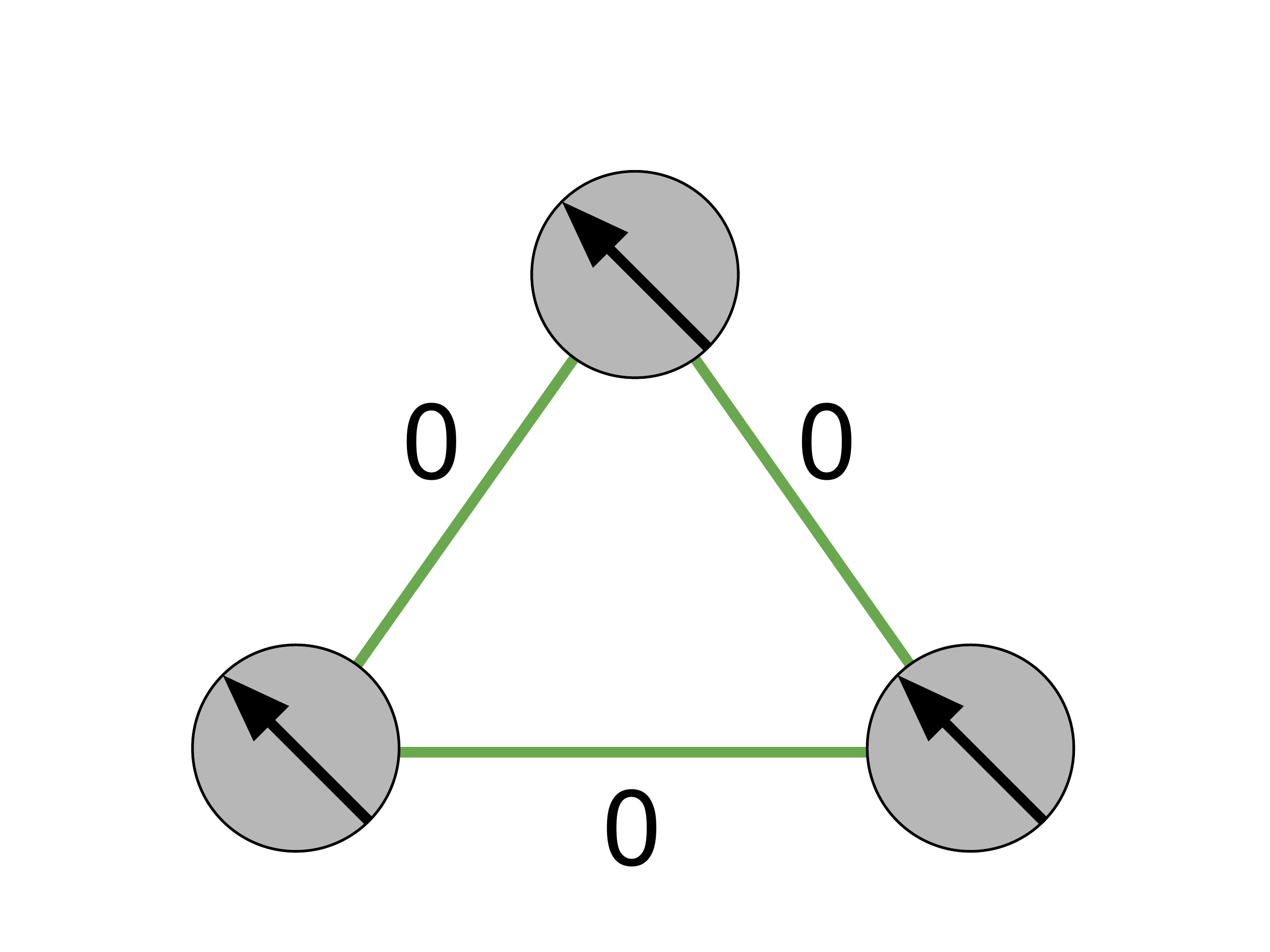}
        \end{subfigure}%
\caption{(Color Online) Modes of oscillation in a triangle of 3 Brusselators. Phase differences between Brusselators are marked. Diagonal lines represent a relative phase.}

\label{fig:cart3}

\end{figure}

A triangular ring is the smallest to exhibit geometric frustration:   in the large limit cycle all Brusselators prefer to be exactly $\pi$ out of phase with their nearest neighbors, which is impossible in this geometry. In contrast to the behavior of  phase-coupled Kuramoto oscillators\cite{Zahera_Mike}, numerical solutions to the mean Field equations show that the Brusselators do not relieve their frustration by adjusting the phase difference to be $2\pi/3$. Instead, the $3-$ring adjusts to the frustration by maximizing the number of $\pi$ out of phase interactions:  synchronizing two of the Brusselators in phase and one Brusselator out of phase with the other two as shown in the schematic in Figure \ref{fig:cart3}.  Additionally,  if the coupling is weak enough the Brusselators are also observed to oscillate in phase. { In experiments on hexagonal arrays of  oscillators\cite{toiya_fraden} with the BZ chemistry, two patterns are observed. At low diffusive coupling the $2\pi/3$ pattern appears, and at strong coupling a $s0\pi$\cite{PNAS_Nate} pattern is observed,  which also maximizes the number of $\pi$ out of phase oscillators. In this pattern however the central droplet stops oscillating while the surrounding six droplets oscillate $\pi$ out of phase.}

%   {\color{red} Figures are now where i had meant to put them, this may or may not be optimal.   I see that the hysteresis plot is there but there is no discussion in the text.  Need to fix that also, do we want to talk about histerisis? we can try and frame it like the rest of the paper, look at where linear stability says it should stop oscillating and then show that it doesn't hold?.}

%Pictures

\subsection{Four Brusselators - Multi Stability}
Expanding the ring to 4 Brusselators eliminates the geometrical frustration associated with odd numbers of oscillators that prefer to be $\pi$ out of phase.   Numerical results, however,  reveal a complex landscape of attractors with multiple modes of oscillations occurring at a fixed set of model parameters.   

Mode 1:  As one would expect from the results of the 2 Brusselator system there is a mode where all Brusselators are out of phase with their neighbors.

\begin{figure}[hbtp]
        
        \begin{subfigure}[b]{0.45\linewidth}
                \includegraphics[width=\linewidth]{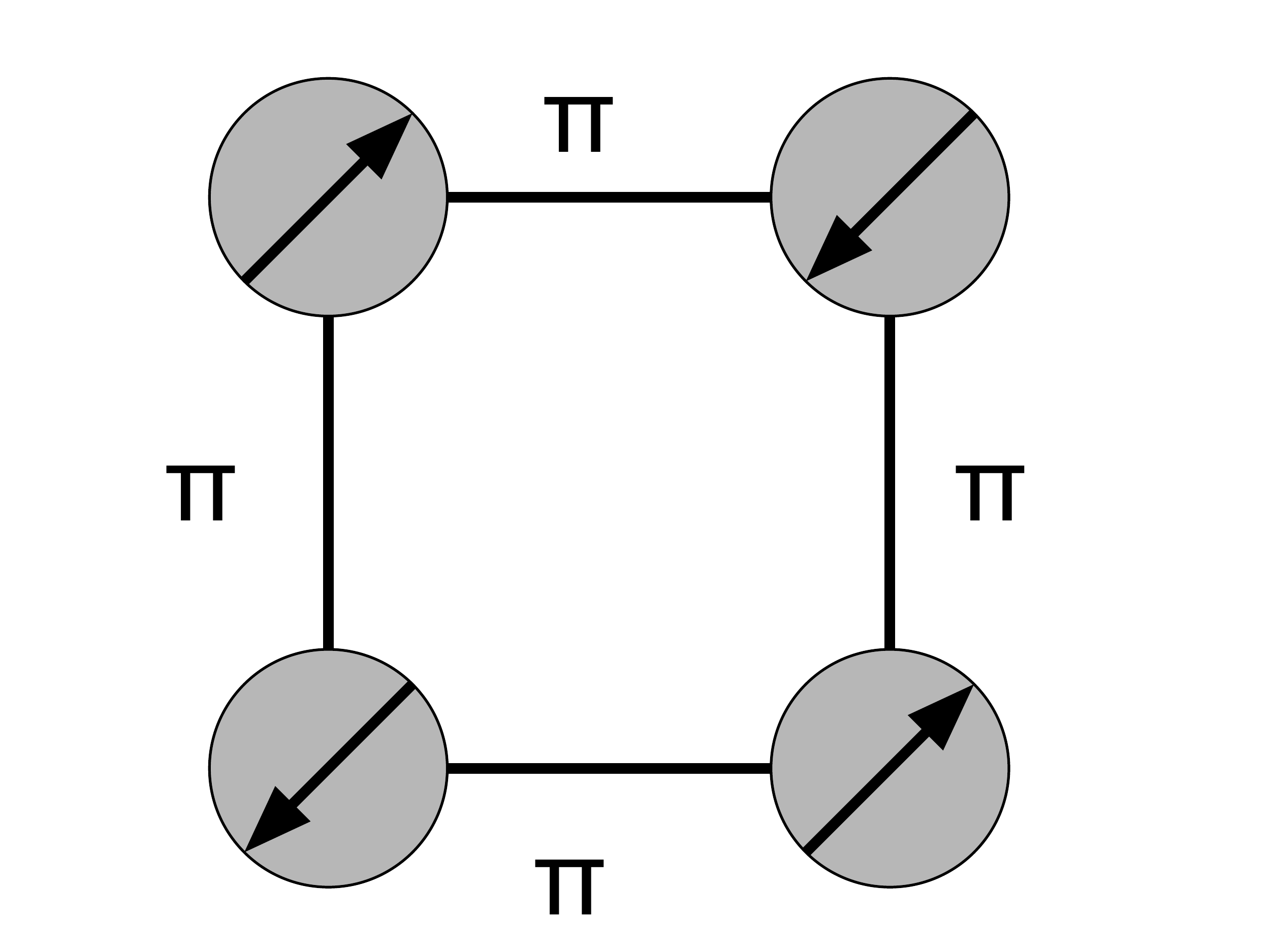}
                \caption{Mode 1}
                \label{fig:4cart1}
        \end{subfigure}%
        ~ %add desired spacing between images, e. g. ~, \quad, \qquad etc.
        \begin{subfigure}[b]{.45\linewidth}
                \includegraphics[width=\linewidth]{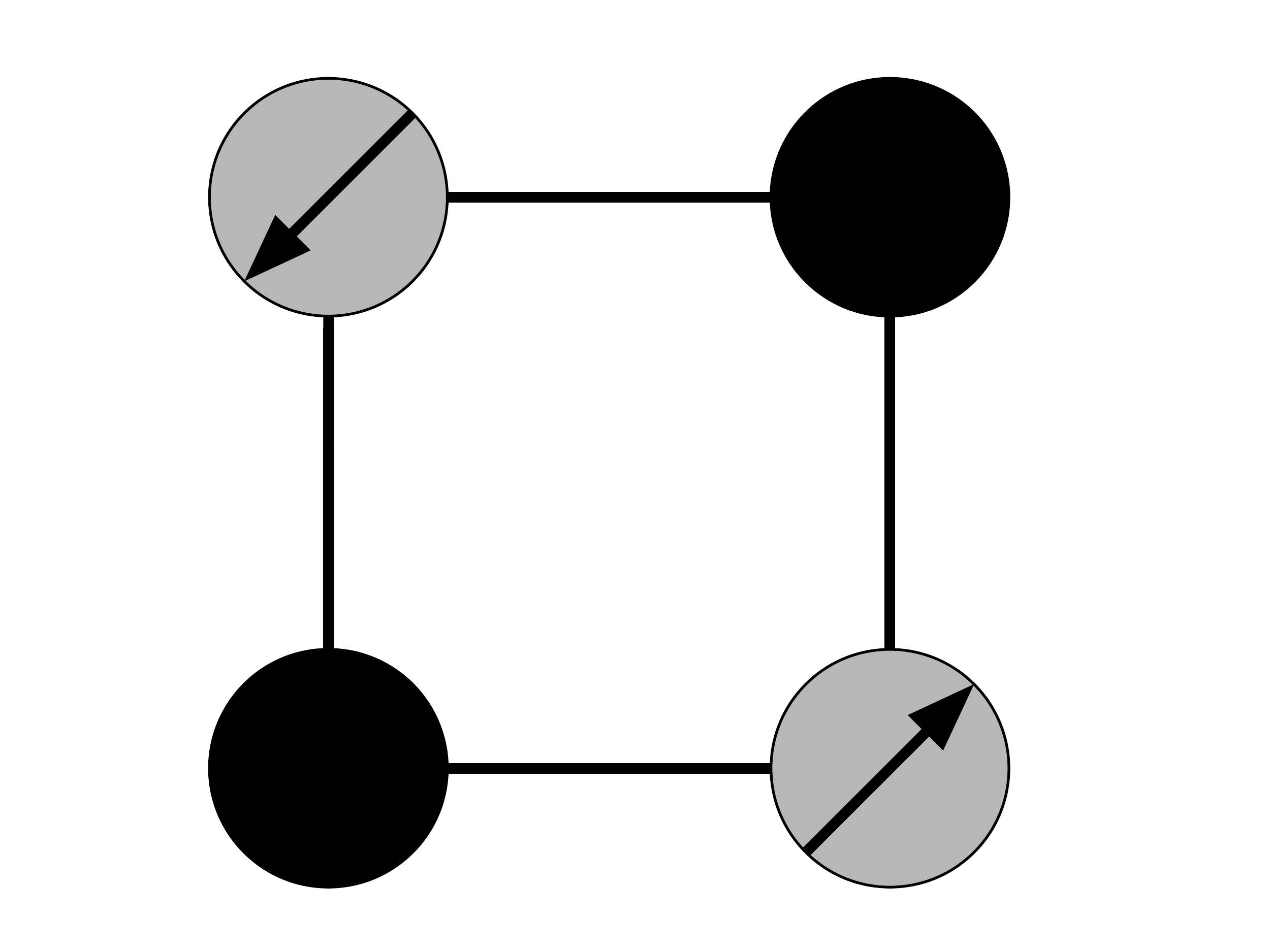}
                \caption{Mode 2}
                \label{fig:4cart2}
        \end{subfigure}
        ~ %add desired spacing between images, e. g. ~, \quad, \qquad etc.
          \\
        \begin{subfigure}[b]{0.45\linewidth}
                \includegraphics[width=\linewidth]{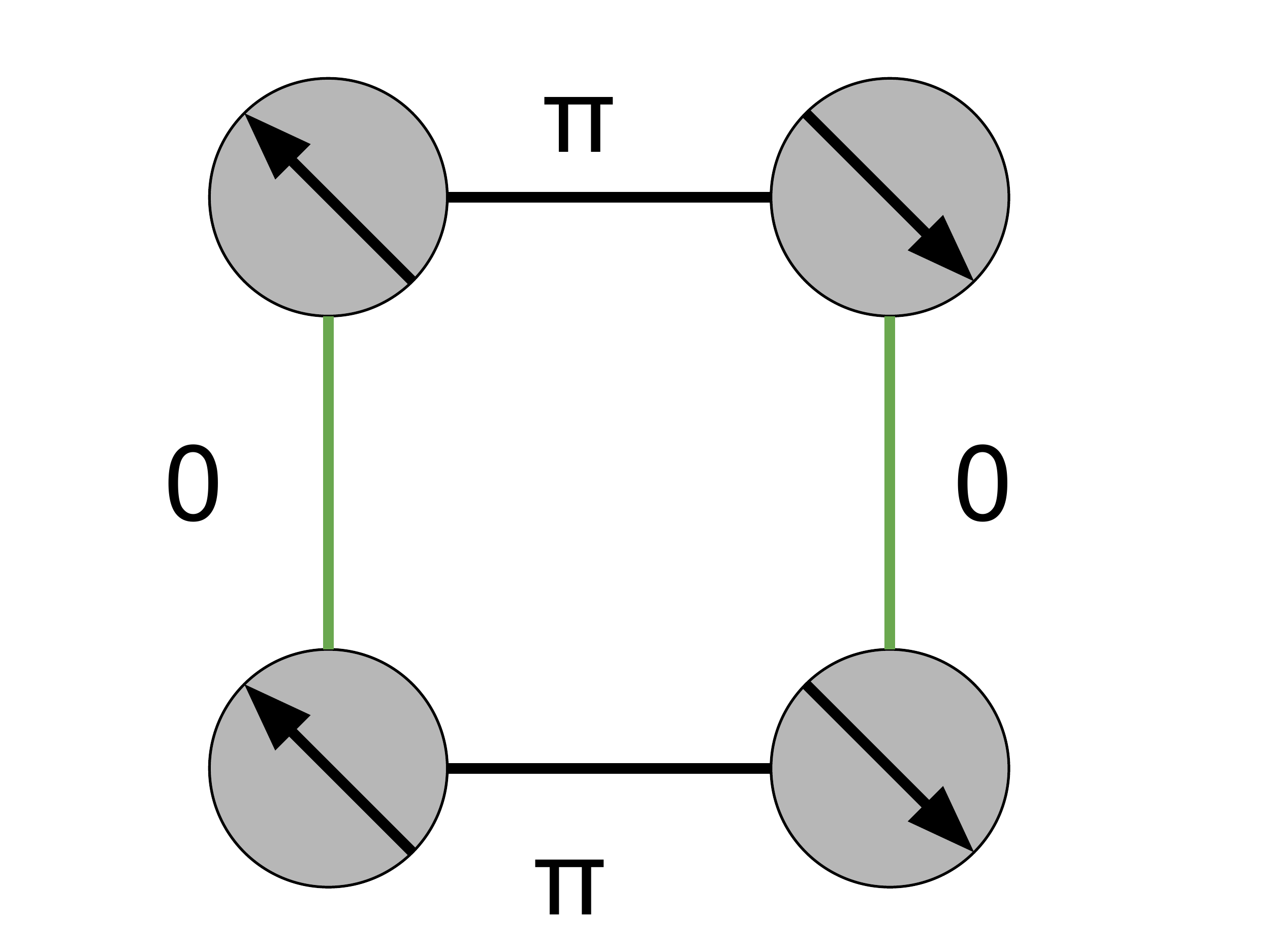}
                \caption{Mode 3}
                \label{fig:4cart3}
        \end{subfigure}
        \begin{subfigure}[b]{0.45\linewidth}
                \includegraphics[width=\linewidth]{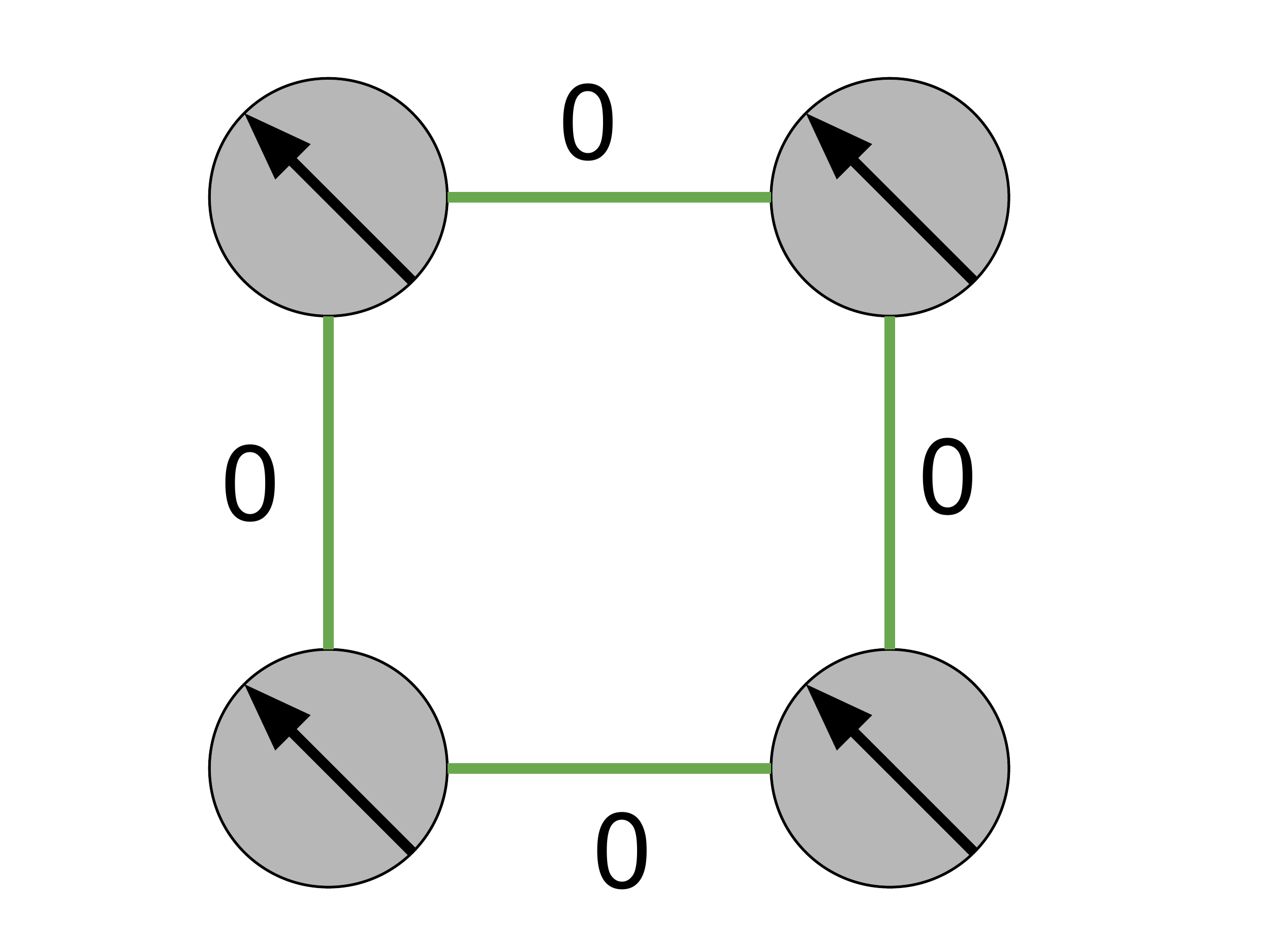}
                \caption{Mode 4}
                \label{fig:4cart4}
        \end{subfigure}
        \caption{(Color Online) Modes of oscillation for a ring of 4 Brusselators. Diagonal lines represent a relative phase. }
        \label{fig:4brussmodes}
        %\newline
        \end{figure}

Mode 2: The first additional mode is one where each Brusselator has one in phase neighbor and one out of phase neighbor, this leads to the sides of square oscillating out of phase with each other as can be seen in the cartoon in Figure \ref{fig:4cart2}.

Mode 3: The next mode of oscillation observed is characterized by two Brusselators that have stopped oscillating on opposite corners complimented by two out of phase oscillating Brusselators on the remaining corners as can be seen in \ref{fig:4cart3}.

Mode 4: Finally there exists a mode where all four Brusselators oscillate in phase.

There exist regions in $\delta-d$ space where  the Brusselators exhibits multistability of these modes. For example at $\delta=1$, $d=0.02$ modes 1, 2 and 3 are all stable periodic solutions accessible from different initial conditions.
\subsection{Five Brusselators - Frustrated patterns}
%\subsubsection{Low $\delta$}
The geometry of this odd numbered ring frustrates the system, as for the three-oscillator system since it is impossible for all the Brusselators to be exactly $\pi$ out of phase with their neighbors. Unlike the $3$-ring, however, we encounter an explosion of synchronized patterns for the $5$-ring.   The oscillating patterns can be characterized by five distinct  relations as shown in Table \ref{tab:5bruss}\\

\begin{figure*}[htp]

\begin{subfigure}[htp]{.4\linewidth}
\caption{}
\includegraphics[width=.9\linewidth]{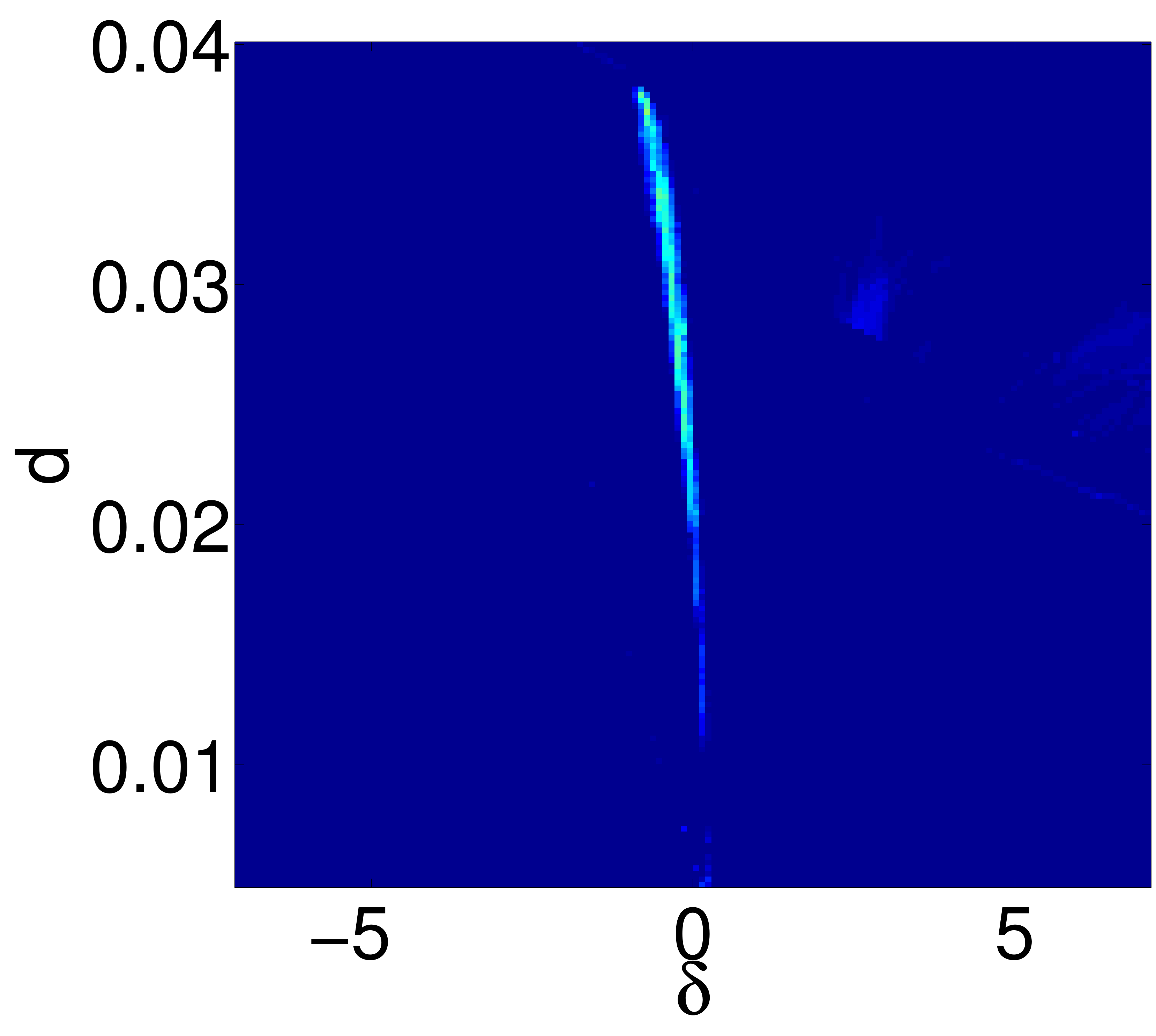}
%Probability of falling into mode 1 in $\delta - d$ space for c=50. Red is high probability blue is low probability}
\label{fig:mode1}
\end{subfigure}
\quad
\begin{subfigure}[htp]{.4\linewidth}
\caption{}
\includegraphics[width=.9\linewidth]{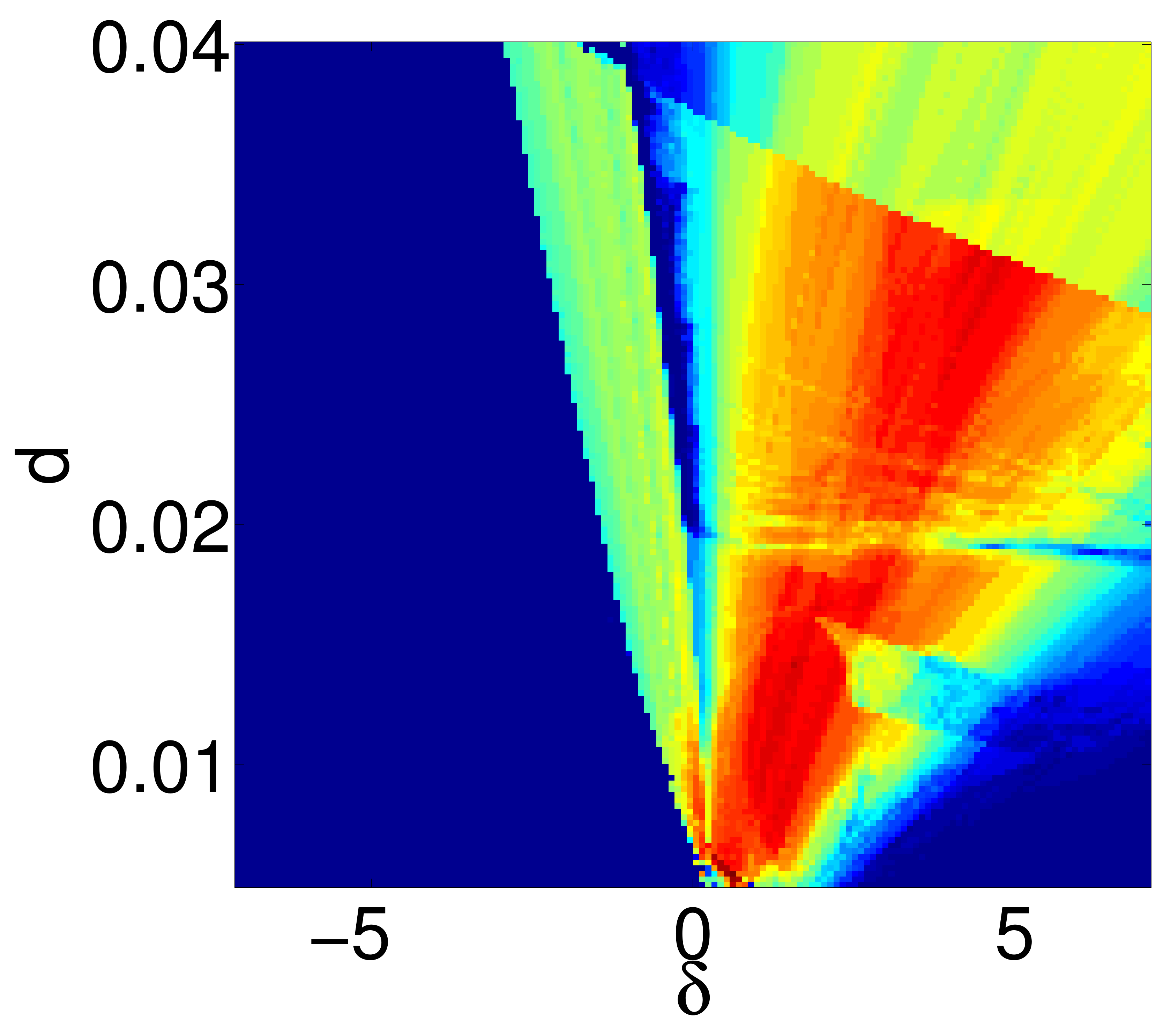}
%Probability of falling into mode 2 in $\delta - d$ space for c=50. Red is high probability blue is low probability}
\label{fig:mode2}
\end{subfigure}

\begin{subfigure}[htp]{.4\linewidth}
\caption{}
\includegraphics[width=.9\linewidth]{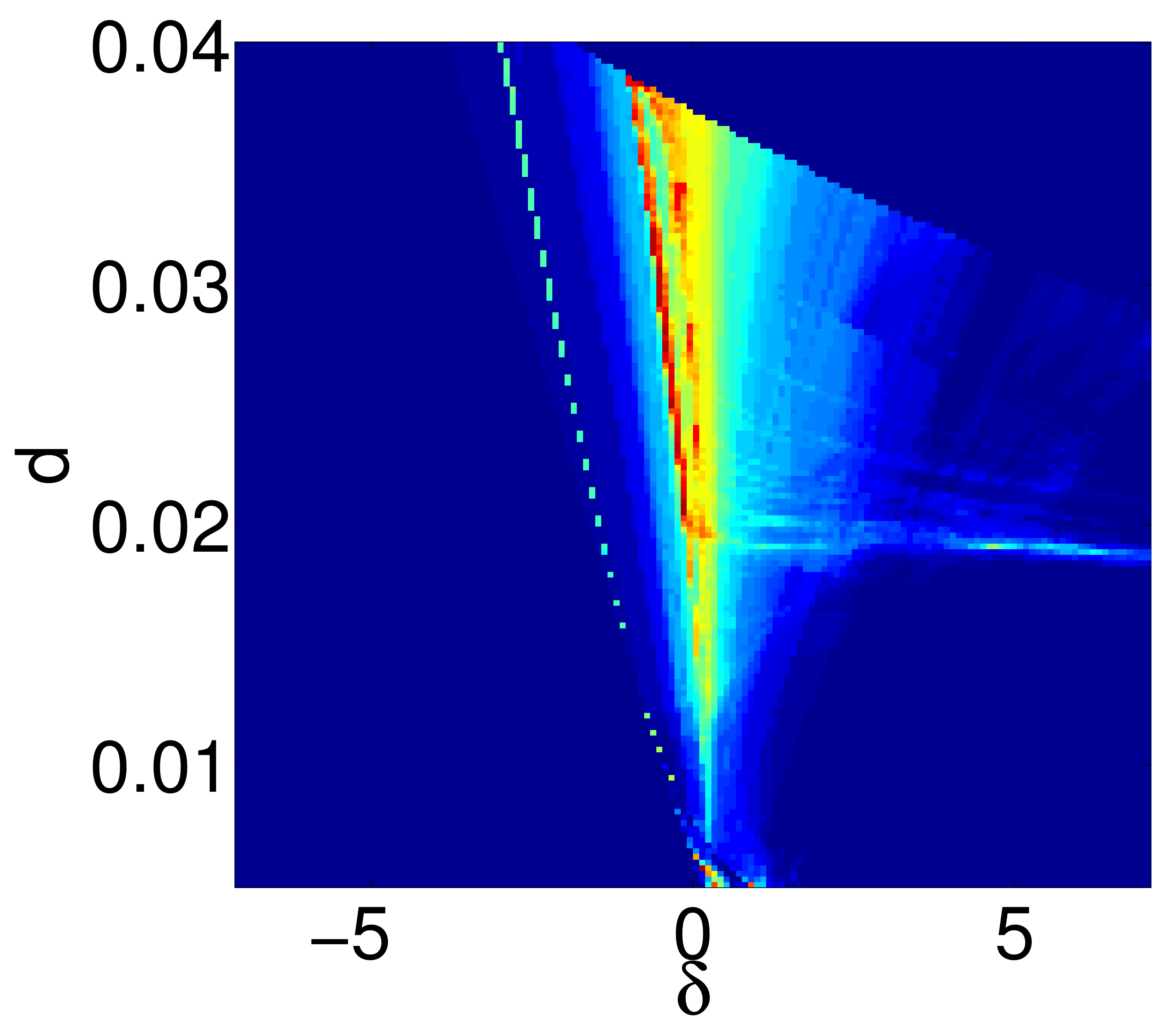}
%Probability of falling into mode 3 in $\delta - d$ space for c=50. Red is high probability blue is low probability}
\label{fig:mode3}
\end{subfigure}
\begin{subfigure}[htp]{.4\linewidth}
\caption{}
\includegraphics[width=.9\linewidth]{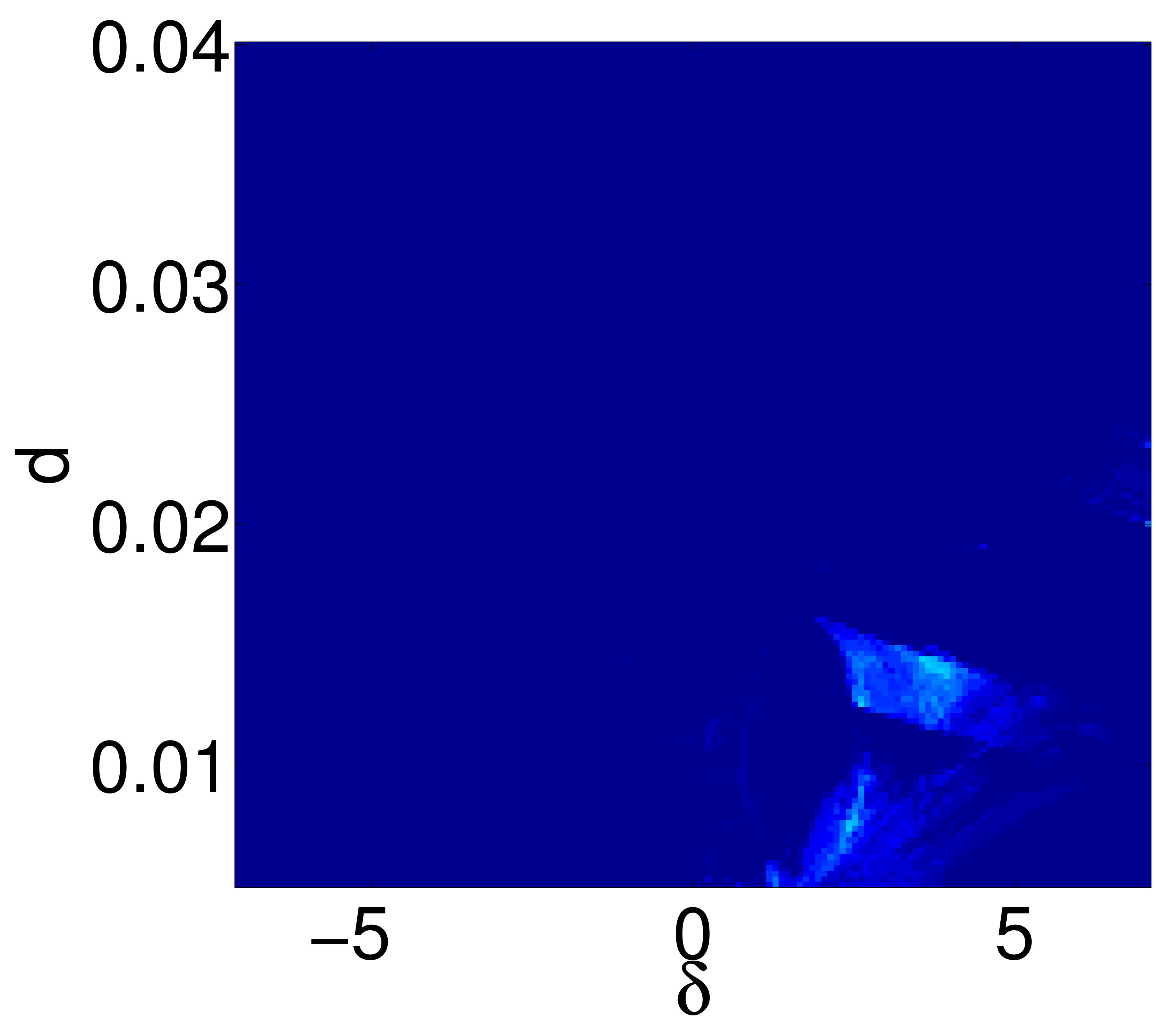}
%Probability of falling into mode 4 in $\delta - d$ space for c=50. Red is high probability blue is low probability}
\label{fig:mode4}
\end{subfigure}

\begin{subfigure}[htp]{.4\linewidth}
\caption{}
\includegraphics[width=.9\linewidth]{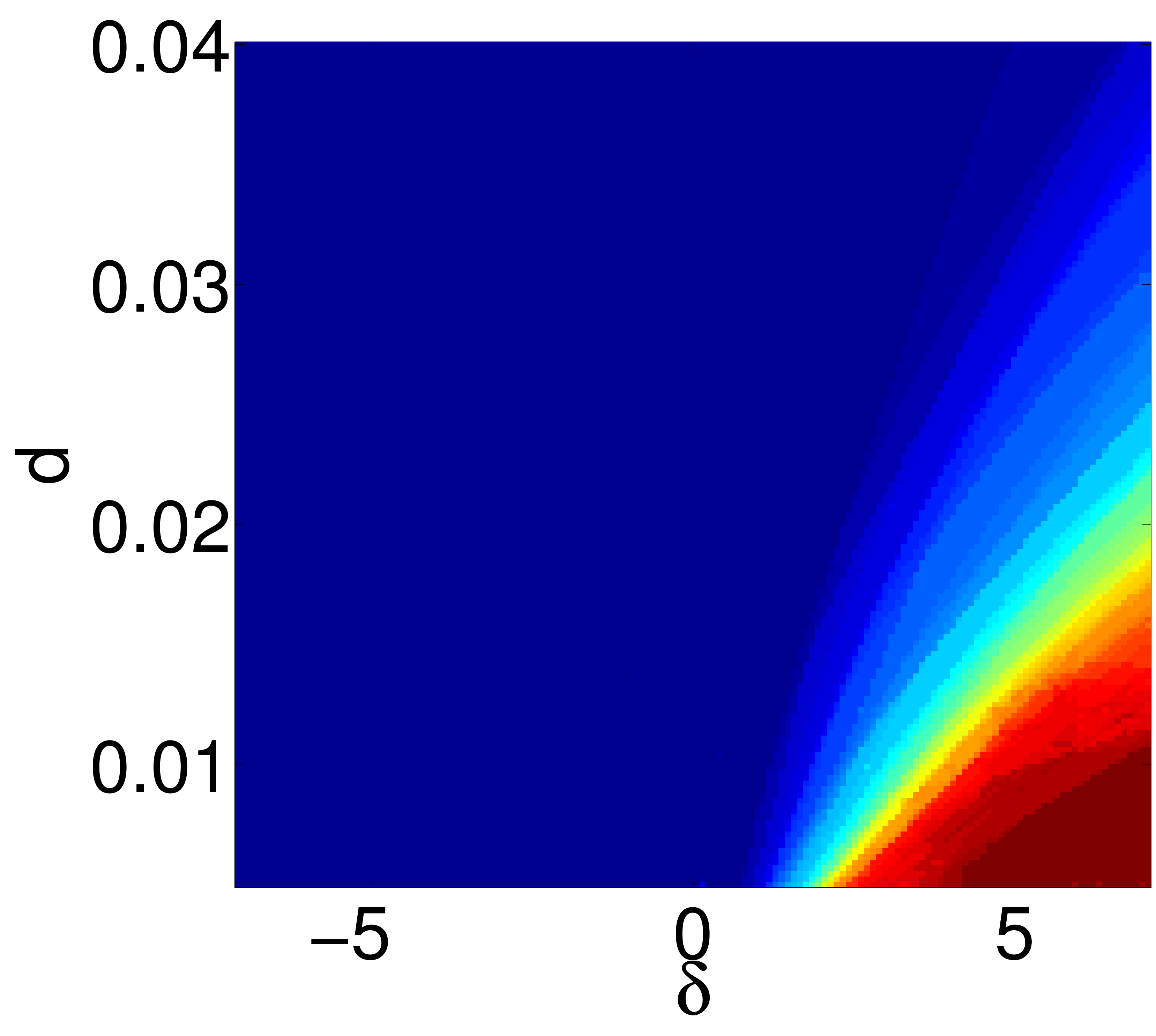}
%Probability of falling into mode 5 in $\delta - d$ space for c=50. Red is high probability blue is low probability}
\label{fig:mode5}
\end{subfigure}
\begin{subfigure}[htp]{.4\linewidth}
\caption{}
\includegraphics[width=.9\linewidth]{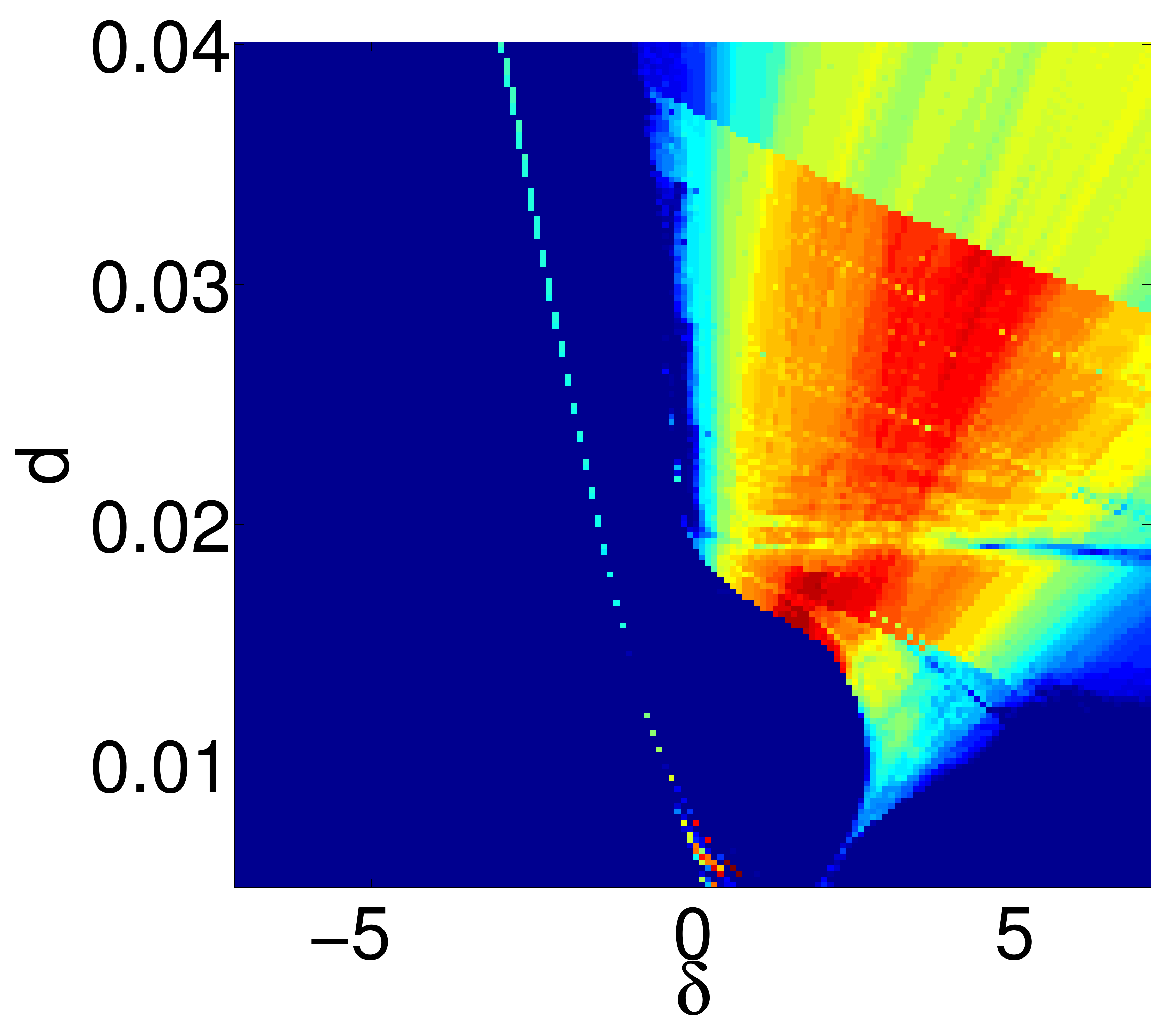}
%Probability of Brusselators having different frequencies in $\delta - d$ space for c=50. Red is high probability blue is low probability}
\label{fig:mode6}
\end{subfigure}
\caption{(Color Online) Probability heat maps for the modes of oscillations of the $5$-ring.  (a)-(e) correspond to  Modes 1 through 5 in Table \ref{tab:5bruss}: (f) shows the  probability that an initial configuration of identical Brusselators have a final state where they oscillate with different frequencies. Color indicates how often that mode of oscillation was observed in an ensemble of  50 randomly chosen initial conditions. Blue indicates that the mode of oscillation was rarely or never achieved for that value of $d$ and $\delta$, while red indicates that the mode was achieved by a majority of the randomly chosen initial conditions. }
\end{figure*}

\begin{table}[htp]
   
   %\topcaption{Table captions are better up top} % requires the topcapt package
   \begin{tabular}{|p{.1\linewidth}|p{.5\linewidth}|c|} % Column formatting, @{} suppresses leading/trailing space
   \hline
  Mode  Number & Description & Schematic\\ \hline
   1 & Nearest neighbors have a phase difference of $4\pi/5$. This creates a star like firing pattern & \includegraphics[width=.3\linewidth]{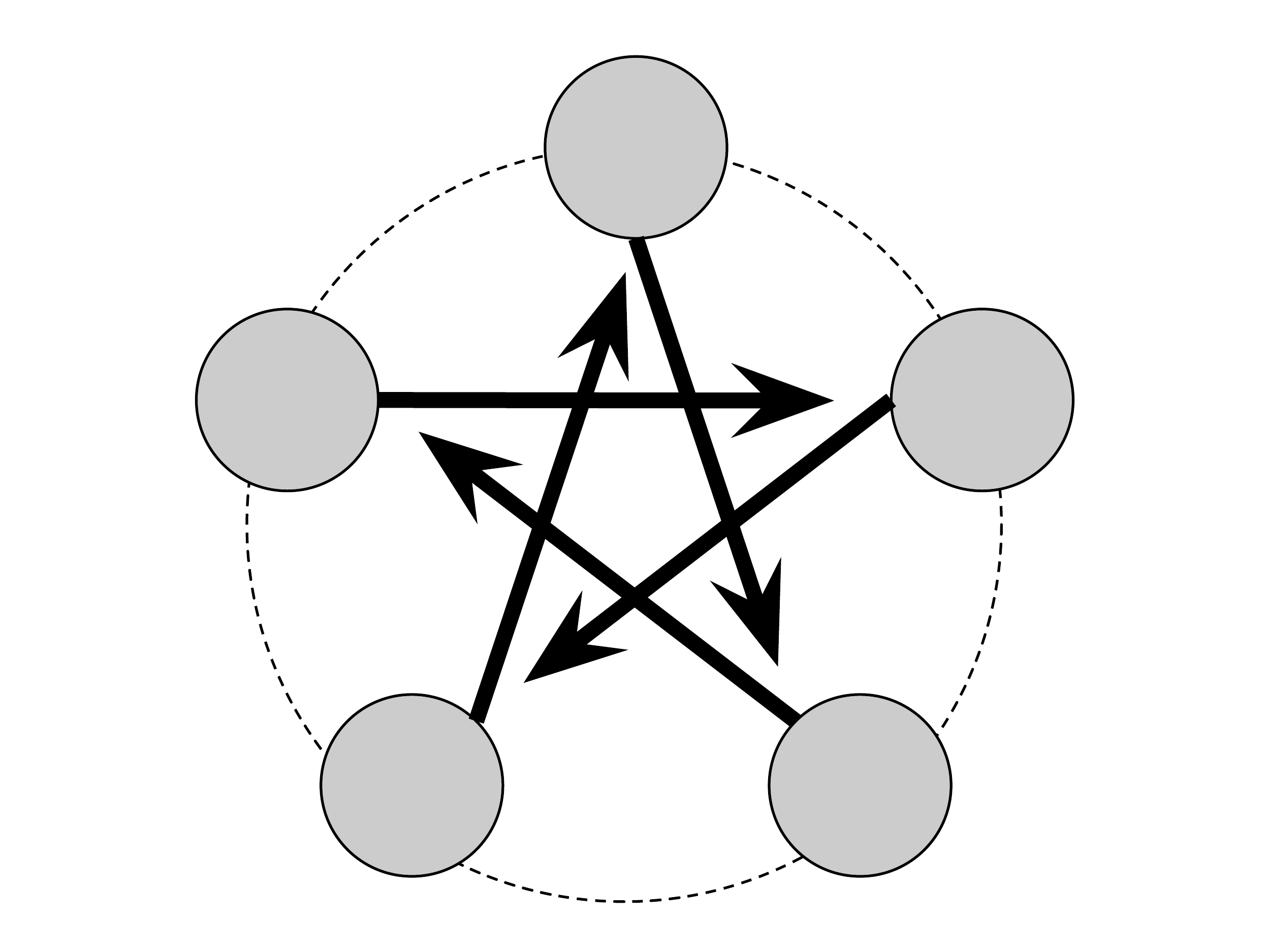}\\\hline
   2 & One pair of nearest neighbors is in phase and  one pair of second nearest neighbors are in phase, leaving one to oscilate by itself. &\includegraphics[width=.3\linewidth]{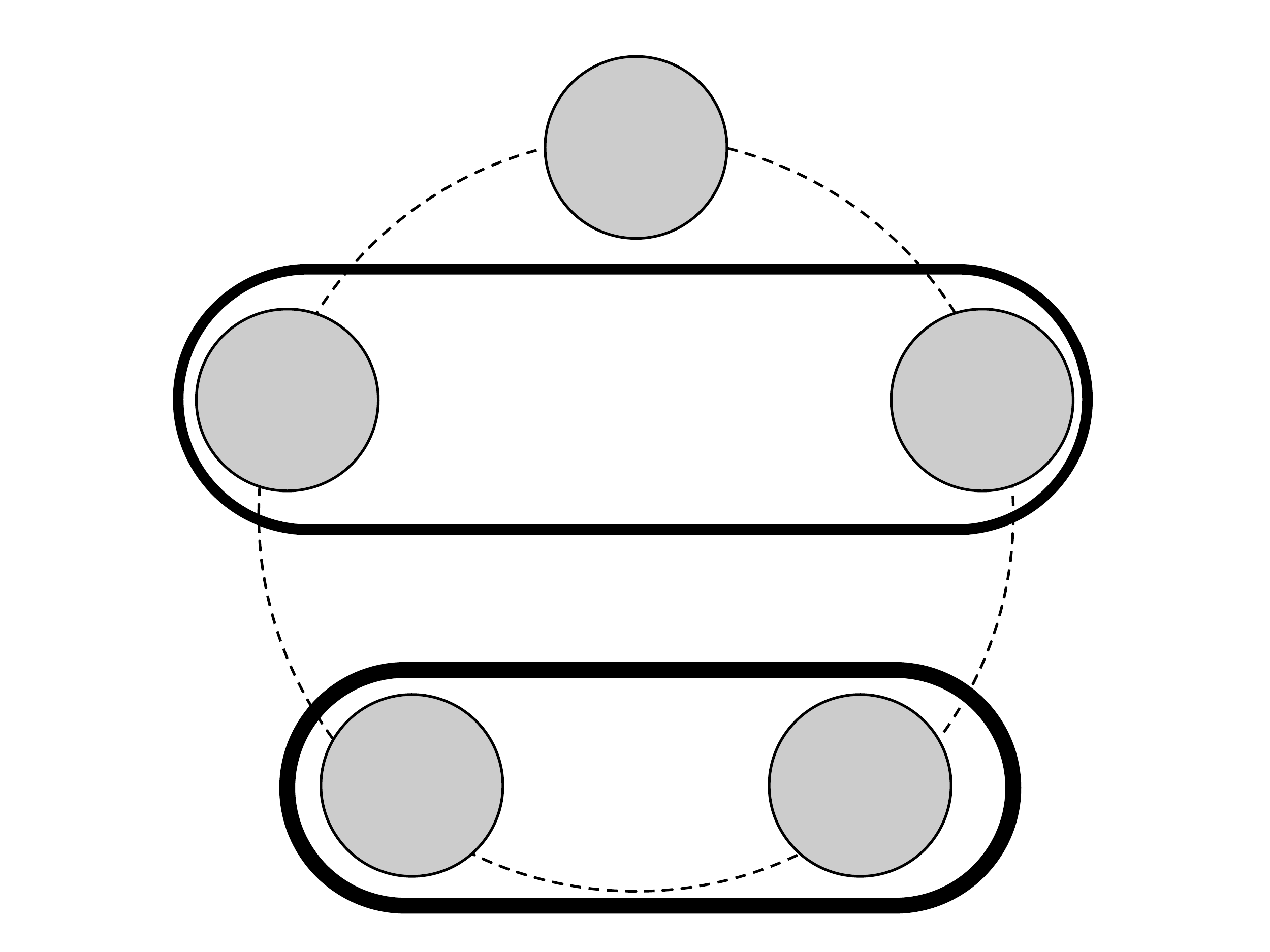}\\\hline
   3 & One pair of second nearest neighbors are in phase this traves around the ring. & \includegraphics[width=.3\linewidth]{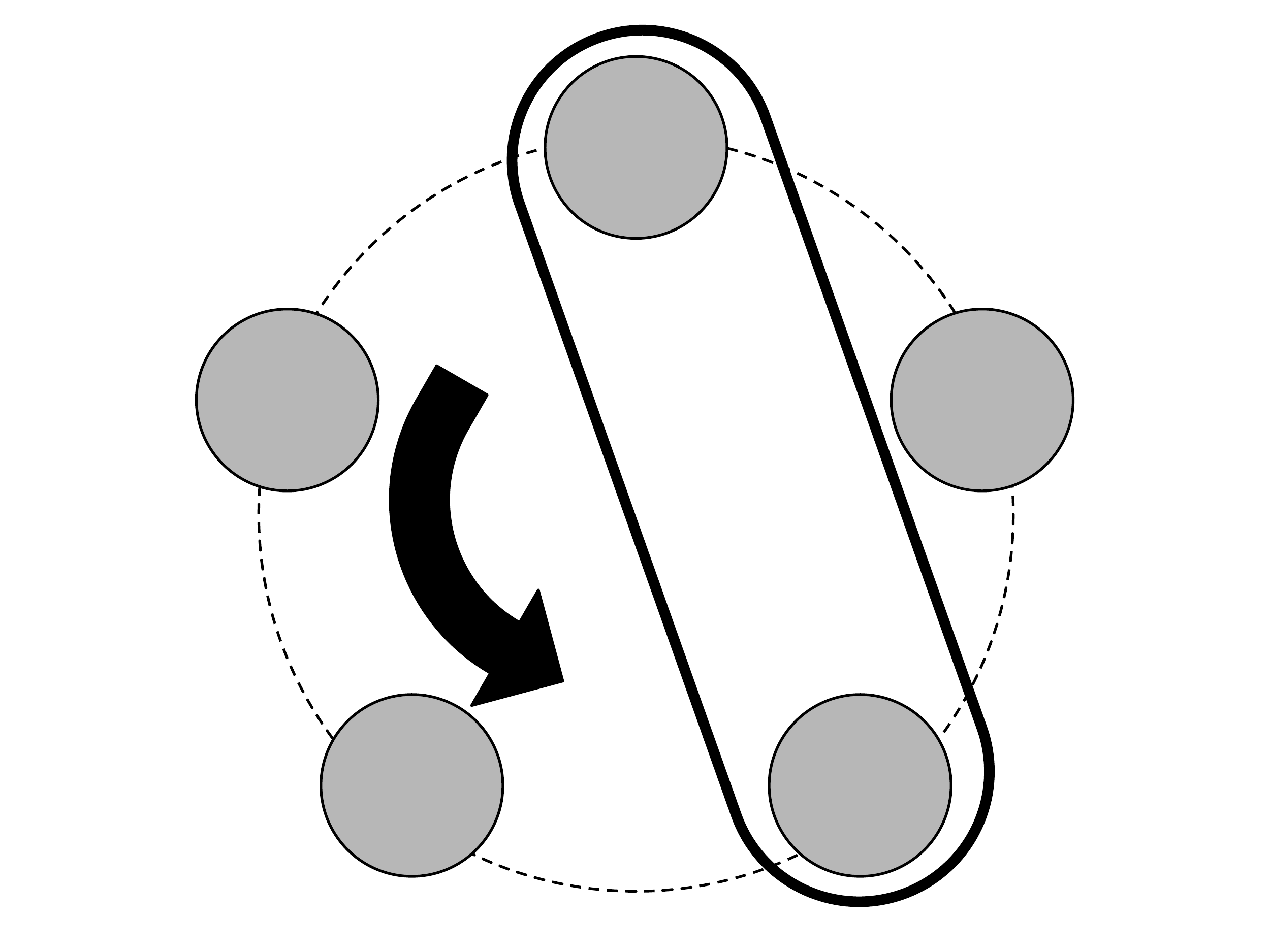}\\\hline
   4 & One pair of second nearest neighbors are in phase and the the remaining 3 are in phase.  & \includegraphics[width=.3\linewidth]{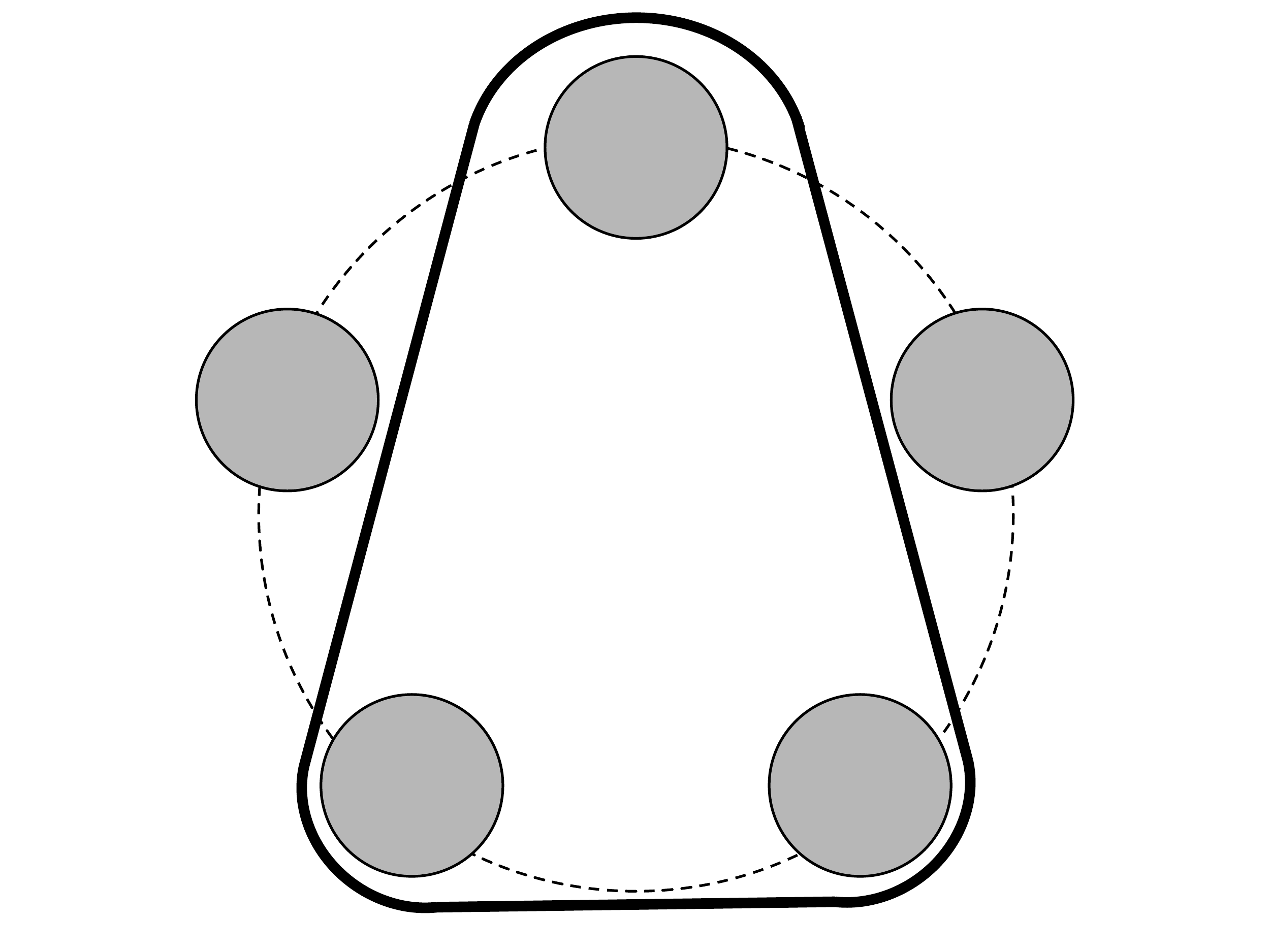}\\\hline
   5 & All are in phase & \includegraphics[width=.3\linewidth]{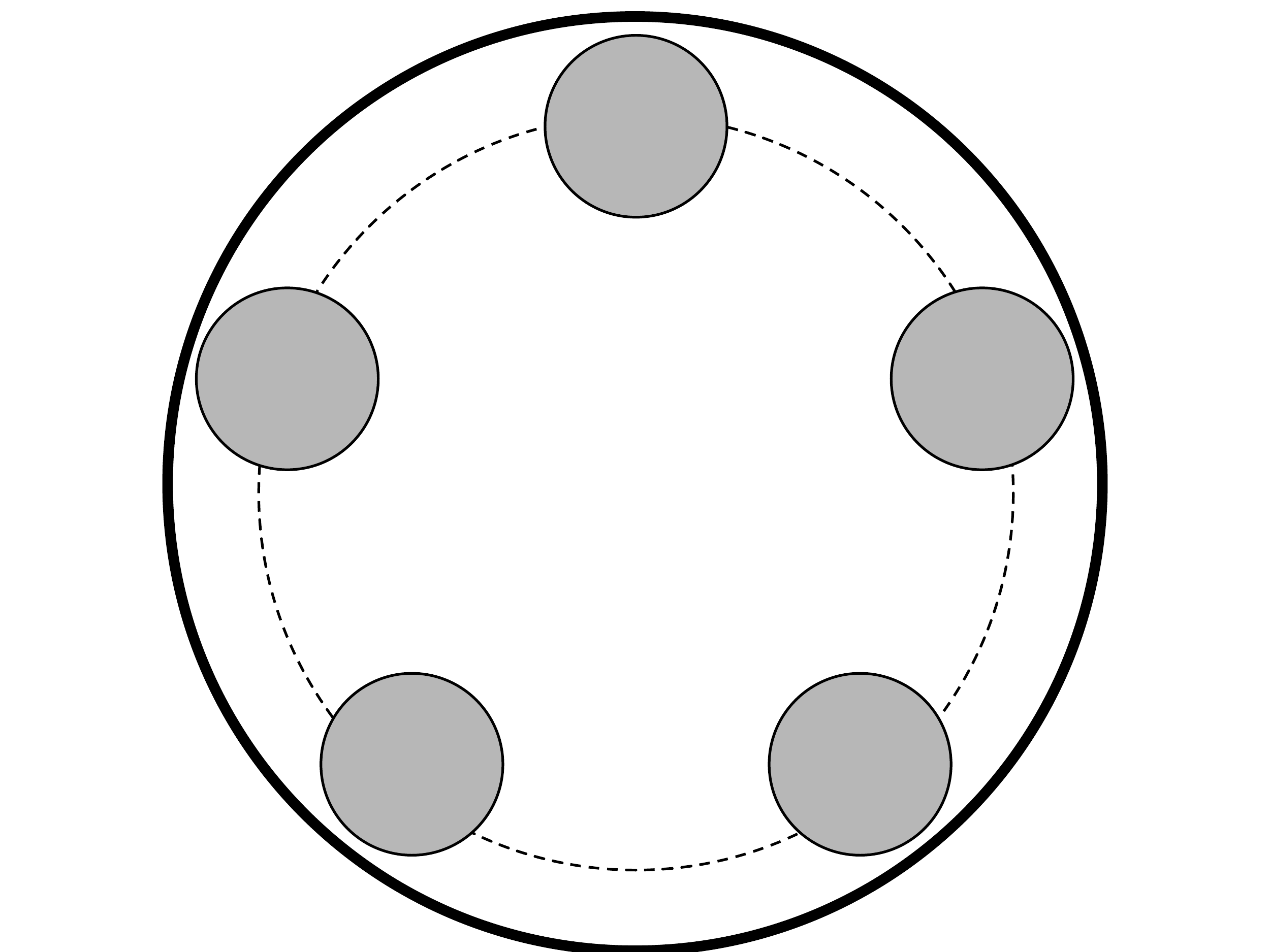}\\\hline
     \end{tabular}
   \caption{Description of phase relationships in observed modes in a ring of five Brusselators}
   \label{tab:5bruss}
\end{table}
Starting from different initial conditions these modes occur with varying probability in different parts of the $\delta - d$ phase space as outlined in Fig.~ \ref{fig:mode1}-~\ref{fig:mode5}. %{\color{red} Need phase diagram figure}
{ Furthermore, when in mode 2 the oscillators can exhibit different frequencies from each other. Specifically,  the synchronized patterns are characterized by a frequency ratio of the three groups of oscillators participating in Mode 2 (Table~\ref{tab:5bruss}), but the ratios change discontinuously as $\delta$ is varied at a fixed $d$ in the region of the phase diagram indicated in Fig.~ \ref{fig:mode1}-~\ref{fig:mode5}.}   This behavior is reminiscent of phenomena such as the Devil's staircase encountered in discrete systems\cite{Bak_Devils}.      
%{\color{red}  Can we indicate in the phase diagram the region where the period skipping occurs ? I still need to do this. Will be done monday morning.}   leading to changes in the repeat  frequency  though an infinite number of ratios of integers.
The piecewise linear model discussed in the next section offers insight into the multitude of patterns observed in rings of  the strong relaxation oscillators:  Brusselators at large $c$.

\begin{figure}[htp]
\includegraphics[width=\linewidth]{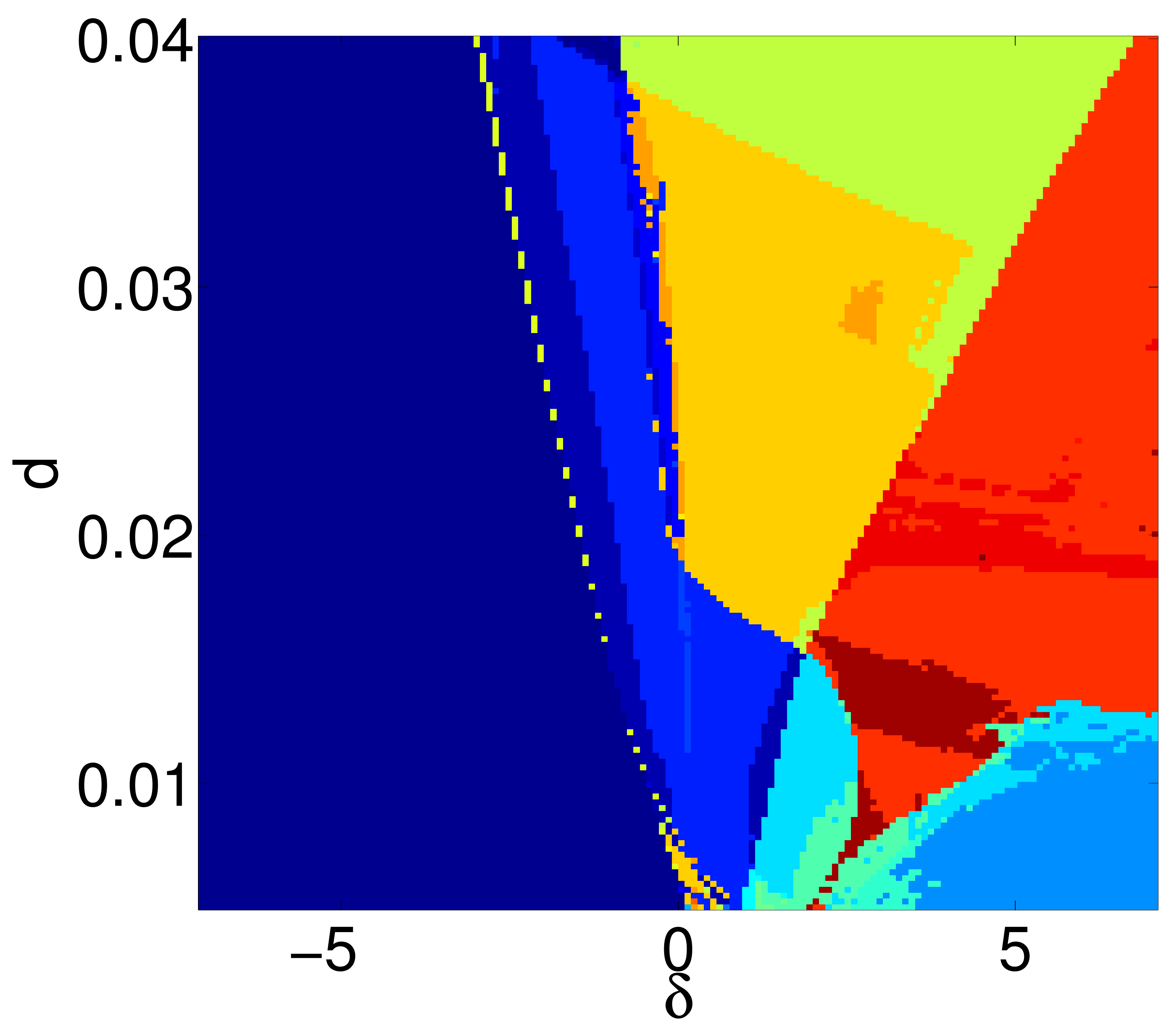}
\caption{(Color Online) Full phase diagram for the $5$-ring illustrating the diversity of patterns.   Each color represents a different combination of possible modes of oscillation achievable from different initial conditions.}
\end{figure}

\section{Piecewise Linear Approximation}
\subsection{Approximating a single Brusselator}
In this section, we analyze the dynamics of $N_r$-rings at large $c$ using a piecewise linear approximation, PLA,   to gain some insight into the multitude of out-of phase synchronized patterns. Specifically, we are trying to understand the origin of the out of phase oscillations and how that gives rise to the multistability of modes. The PLA approximation has been used in  previous studies to analyze strong relaxation oscillators, such as the Van der Pol oscillator\cite{Piecewise} and the Tyson-Fife model \cite{TF}. This has proven to be an effective method for studying relaxation oscillators with different coupling mechanisms including diffusive coupling\cite{Piecewise} and delay coupling\cite{vanDelay}, and in systems with two\cite{Piecewise}, three\cite{RingRelax} and four oscillators\cite{fourvanderpol}.
% why even in a frustrated geometry such as a $3$ or $5$-ring, these non-linear oscillators prefer to maintain  a  $\pi$-out-of-phase relationships, whereas pure phase oscillators would synchronize with a phase difference of $2\pi/N$\cite{Zahera_Mike} { We don't actually talk about this here. Should we include a dissuison of 3 brusselators in the PLA?}

We begin our analysis with the  single Brusselator described in terms of the fast-slow variables: $u$ and $v$. 
\begin{align} % requires amsmath; align* for no eq. number
   \frac{1}{c}\dot{u}=&f(u,v)\\
   \dot{v}=&g(u,v)
\end{align}

where
\begin{align} % requires amsmath; align* for no eq. number
   f(u,v)=& (v-u)\left(\frac{1+\delta}{c}+1-u(v-u)\right) \\
   g(u,v)=& 1-v+u
\end{align}
We can see from Fig. \ref{fig:largecphase} that the  large limit cycle consists of segments that closely follow the  the nullclines of  $u$, the fast variable, with  near instantaneous jumps between the branches of the  nullclines. Specifically,  the limit cycle climbs up the line $u=v$ until it falls off at $(u,v)=(u_{max},u_{max})$ where $u_{max}=\frac{(c+\delta+2)^2}{4c}$.{ We determine $u_{max}$ from the maximum of the $x$ nullcline in the $x, y$ representation.} After reaching $u_{max}$, the oscillator jumps to the left  most branch of $v=u+\frac{\delta+1+c}{c}\frac{1}{u}$ and starts following that branch downwards at coordinates $(u,v)=(\frac{1}{2}(u_{max}-\sqrt{u_{max}^2-4(\delta+1+c)/c}),u_{max})$ until it falls off this branch at the minimum of this nullcline at $(u,v)=(\sqrt{(\delta+1+c)/c},2\sqrt{\delta+1+c/c})$.

\begin{figure}[H] %  figure placement: here, top, bottom, or page
%   \centering
\vspace{2pt}
   \includegraphics[width=.9\linewidth]{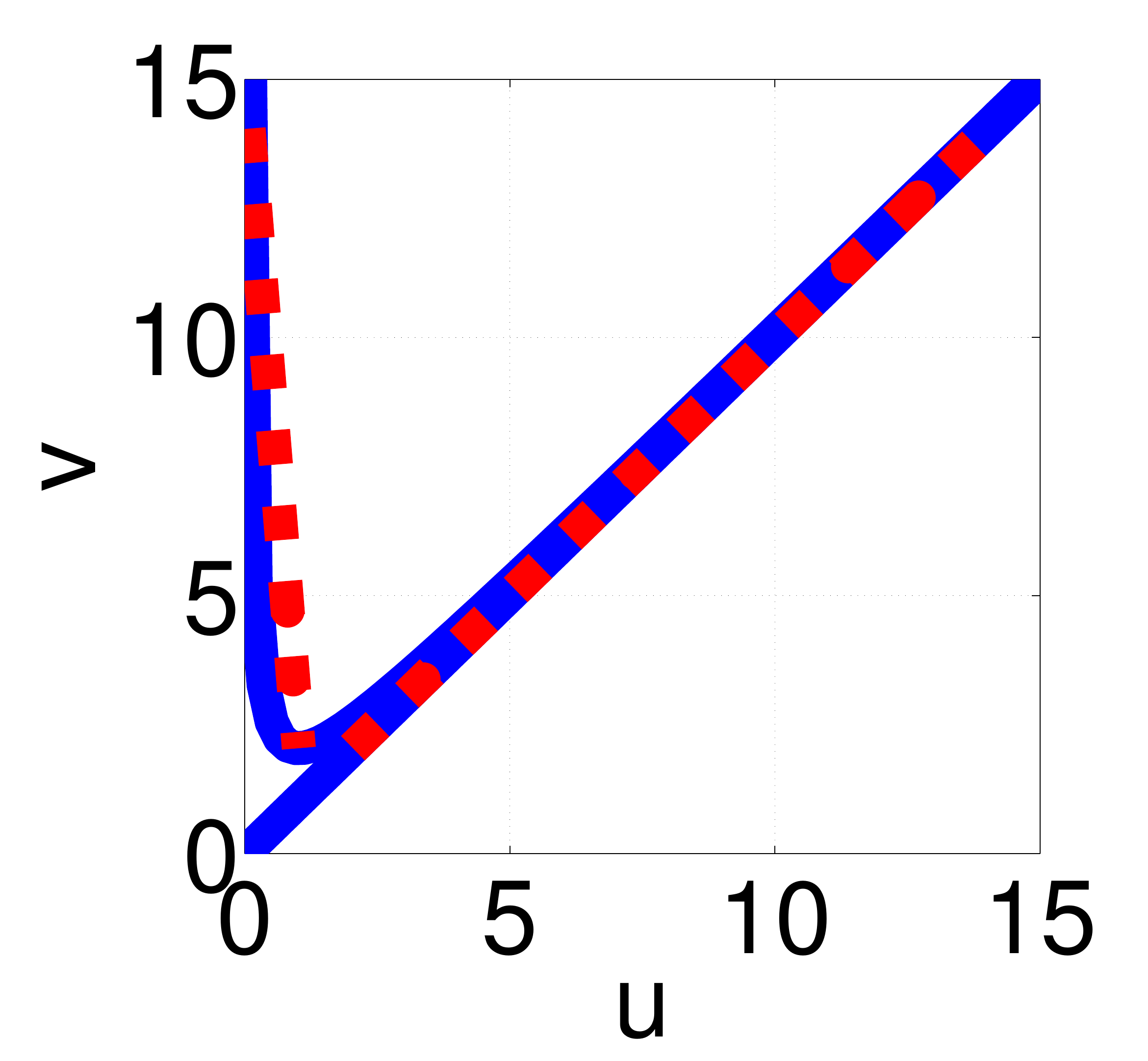} 
   \caption{(Color) Null clines of $u$ from the full model (blue), and from the piecewise linear approximation (red)}
   \label{fig:nullPLA}
\end{figure}
To capture the essence of these dynamics,  we approximate the trajectory by restricting it to the nullclines of $u$ connected by {\it instantaneous} horizontal jumps: an approximation that becomes exact in the limit of $c \rightarrow \infty$. Further,  to make the model analytically solvable we approximate the nullclines of $u$ by linear functions through the replacement:  $f(u,v) \rightarrow f_{pla}(u,v)=v-\Phi(u)$.
Using the instantaneous-jump approximation, we obtain $v$ on $v=\Phi(u)$ by solving the first order linear equation $\dot{v}=1-v+\Phi^{-1}(v)$.
\begin{widetext}
\begin{equation}
\Phi(u) \! \! = \! \! \left\{ \! \!
\begin{array}{lr}
-u\frac{(u_{max}-2\alpha)}{\alpha-\beta}+\alpha\frac{ u_{max}-2\beta}{\alpha-\beta}\! \! &\!\!\beta\!<u\!<\alpha\\
u & 2\alpha<u<u_{max}
\end{array}
\right .
\end{equation}
\end{widetext}
Where we have defined  $\beta=\frac{1}{2}(u_{max}-\sqrt{u_{max}^2-4b/c})$ and $\alpha=\sqrt{\frac{\delta+1+c}{c}}$. We solve the equation in the intervals $\beta<u<\alpha$ and $2\alpha<u<u_{max}$ individually and then switching between the solutions in the two intervals when the solutions  reach $(u,v)=(\alpha,2\alpha)$ and $(u,v)=(u_{max},u_{max})$.  
\begin{figure}[htp] %  figure placement: here, top, bottom, or page
	\centering
	\begin{subfigure}[t]{0.45\linewidth}
		   \includegraphics[width=\linewidth]{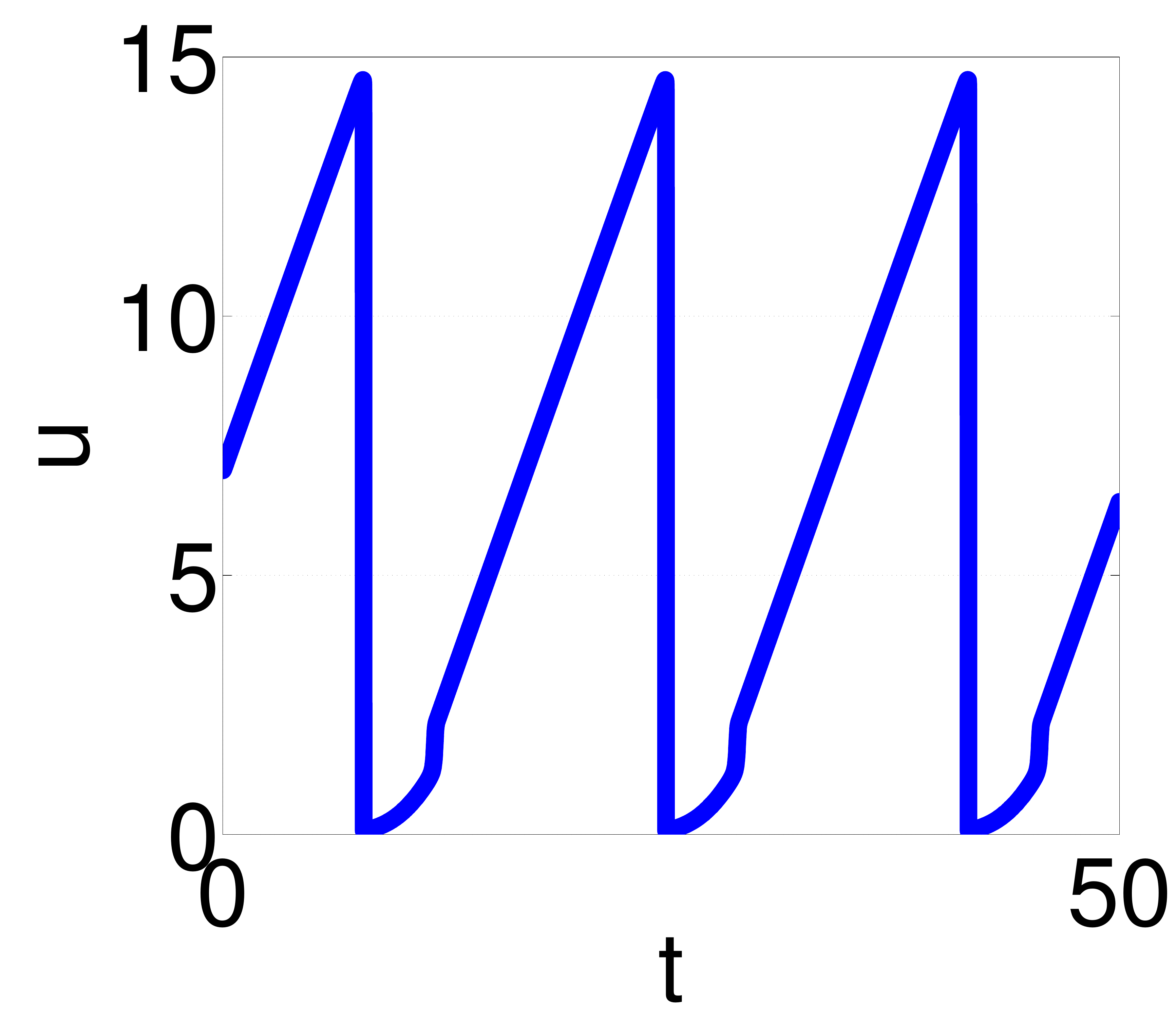} 
   %\caption{{ $u(t)$ from the full model.}}
   \end{subfigure}
   \quad
        \begin{subfigure}[t]{0.45\linewidth}
   \includegraphics[width=\linewidth]{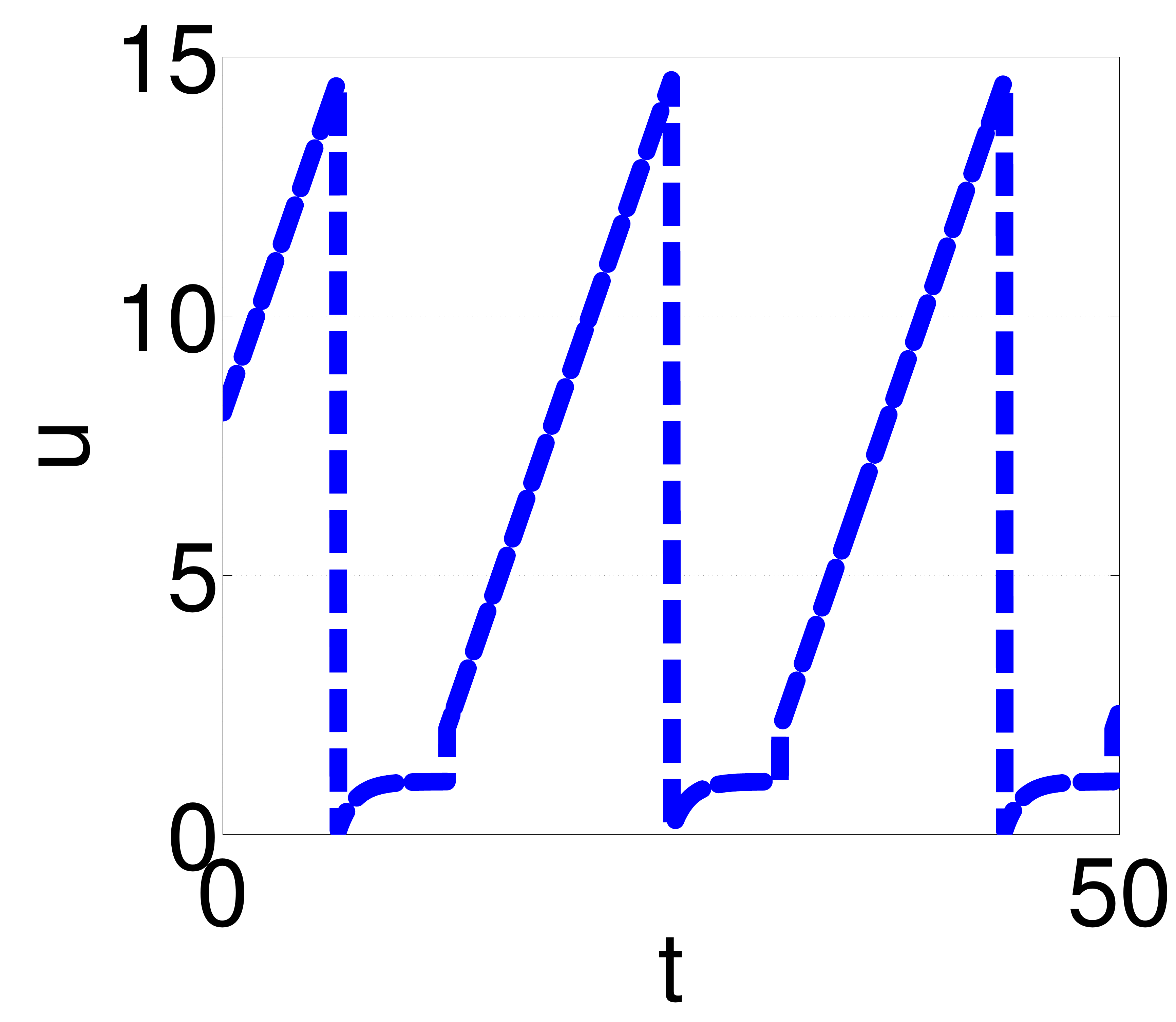} 
   %\caption{{ $u(t)$ from the PLA.}}
   \end{subfigure}
  \caption{Left - Numeric solution to the single Brusselator at c=50. Right - Solution from the PLA for the same parameters.}
   \label{fig:PLAtime}
\end{figure} 

As can be seen from Fig. \ref{fig:nullPLA}, the resulting nullclines of $u$ are linear functions and we are now interested in solving: 
\begin{align} % requires amsmath; align* for no eq. number
   \frac{1}{c}\dot{u}=f_{pla}(u,v)={}& v-\Phi(u) \\
   \dot{v} =g(u,v)={}& 1-v+u\nonumber
\end{align}

In order to examine the dynamics close to the nullclines,  it is more convenient to work in terms of $u$ rather than $v$: $\dv{v}{t}=\pdv{\Phi(u)}{u}\dv{u}{t}=-\frac{(u_{max}-2\alpha)}{\alpha-\beta}\dv{u}{t} \equiv -\eta\dv{u}{t}$ for $\beta<u<\alpha$ and $\dv{v}{t}=\dv{u}{t}$ for $2\alpha<u<u_{max}$. The resulting differential equation for $u$ is: 
\begin{widetext}
\begin{equation}\dot{u}=\left\{
\begin{array}{ll}
-\frac{1}{\eta}[1+u-\Phi(u)]& \beta<u<\alpha\\
1 & 2\alpha<u<u_{max} 
\end{array}
\right .
\end{equation}
\end{widetext}
In the large $c$ regime,  $$u_{max}=\frac{(2+\delta+c)^2}{4c}\approx\frac{c}{4}~,$$ and we know that $\alpha>\beta$ and $u_{max}>2\alpha$.
In addition to the time evolution described above,  the complete dynamics includes instantaneous jumps between the two domains:  $L=(\beta,\alpha)$ $R=(2\alpha,u_{max})$.  As seen from Fig.~\ref{fig:PLAtime},  the PLA captures the functional form of $u(t)$ except for the concavity of L.

\subsection{PLA analysis of two coupled Brusselators} The real advantage of the PLA is that it allows us to explore the attractors and phase diagram of coupled oscillators, and thus provide additional insight into the mechanisms underlying multistability and a complex attractor landscape.   In this section, we explicitly study the system of two diffusively coupled Brusselators using the PLA.    The dynamics close to the nullclines is now generalized to:
\begin{widetext}
\begin{equation}
\dot{u_1}=\left\{
\begin{array}{lr}
-\frac{1}{\eta}[1-\Phi(u_1)+u_1-d(u_1-u_2)]&u_1\in L\\
1 -d(u_1-u_2) & u_1\in R
\end{array}
\right .
\end{equation}
\begin{equation}
\dot{u_2}=\left\{
\begin{array}{lr}
-\frac{1}{\eta}[1-\Phi(u_2)+u_2+d(u_1-u_2)]&u_2\in L\\
1 +d(u_1-u_2) & u_2\in R
\end{array}
\right .
\end{equation}
\end{widetext}
The salient feature that emerges from these equations  is that 
a Brusselator  on the $L$  branch experiences an effective coupling  that is {\it opposite} in sign to the coupling it experiences on branch $R$:   the coupling is repulsive on $L$ and attractive on $R$.  

In the PLA, the original nonlinearity of the equations emerges as jump conditions between two distinct branches of linear dynamics. Another consequence of this partitioning of the dynamics is that the coupling between oscillators changes sign between branches. This is the basic mechanism underlying the complexity in all $N_r$ rings. 

\begin{figure}[htp]
\begin{subfigure}[htp]{.9\linewidth}
\caption{}
		   \includegraphics[width=.9\linewidth]{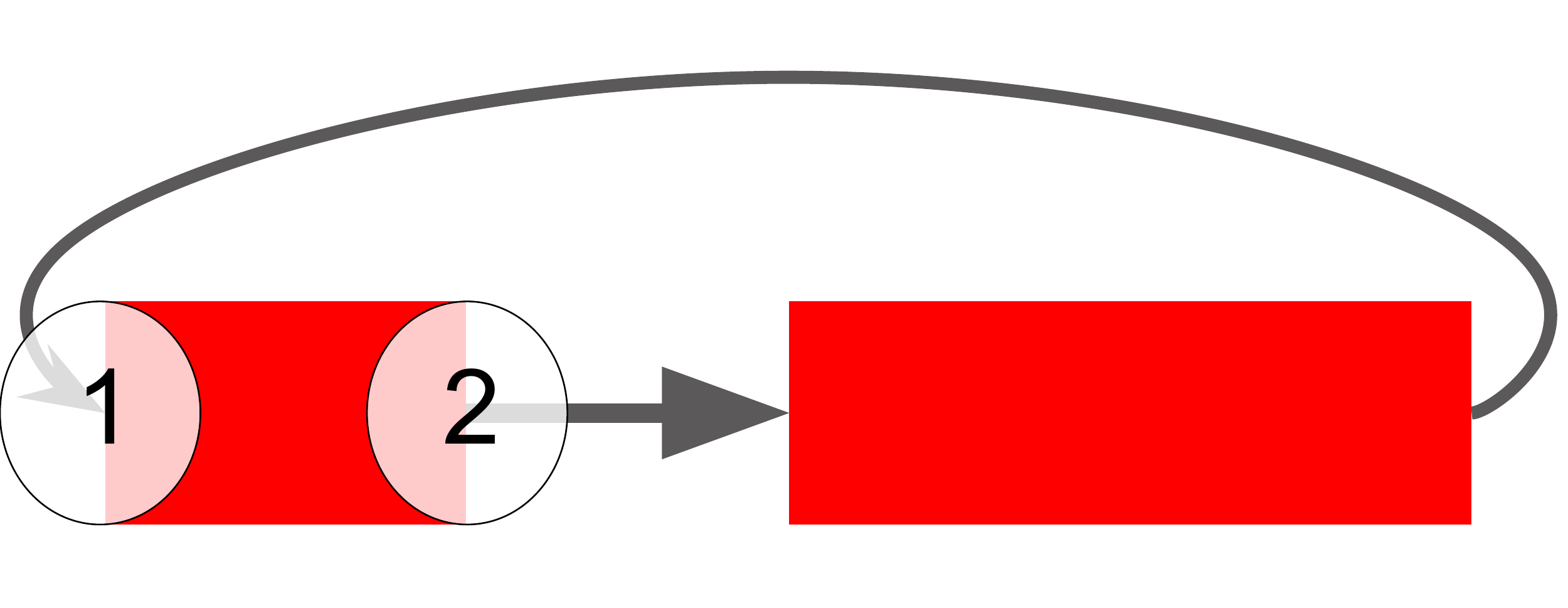} 
		   \label{placart1}
   %\caption{{ $u(t)$ from the full model.}}
   \end{subfigure}
   \begin{subfigure}[hpt]{0.9\linewidth}
   \caption{}
		   \includegraphics[width=.9\linewidth]{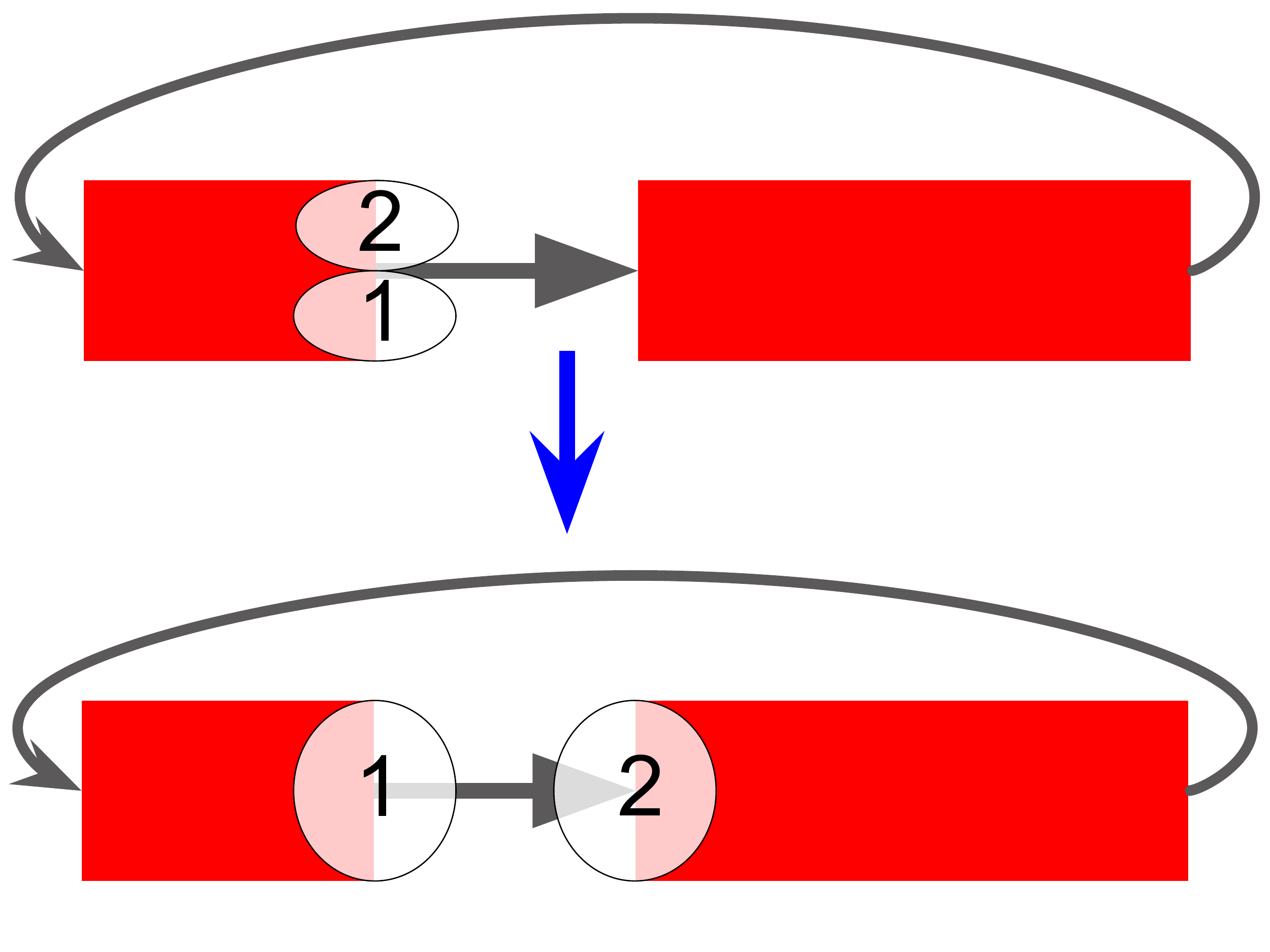} 
		   \label{placart2}
   %\caption{{ $u(t)$ from the full model.}}
   \end{subfigure}
   \begin{subfigure}[htp]{0.9\linewidth}
   \caption{}
		   \includegraphics[width=.9\linewidth]{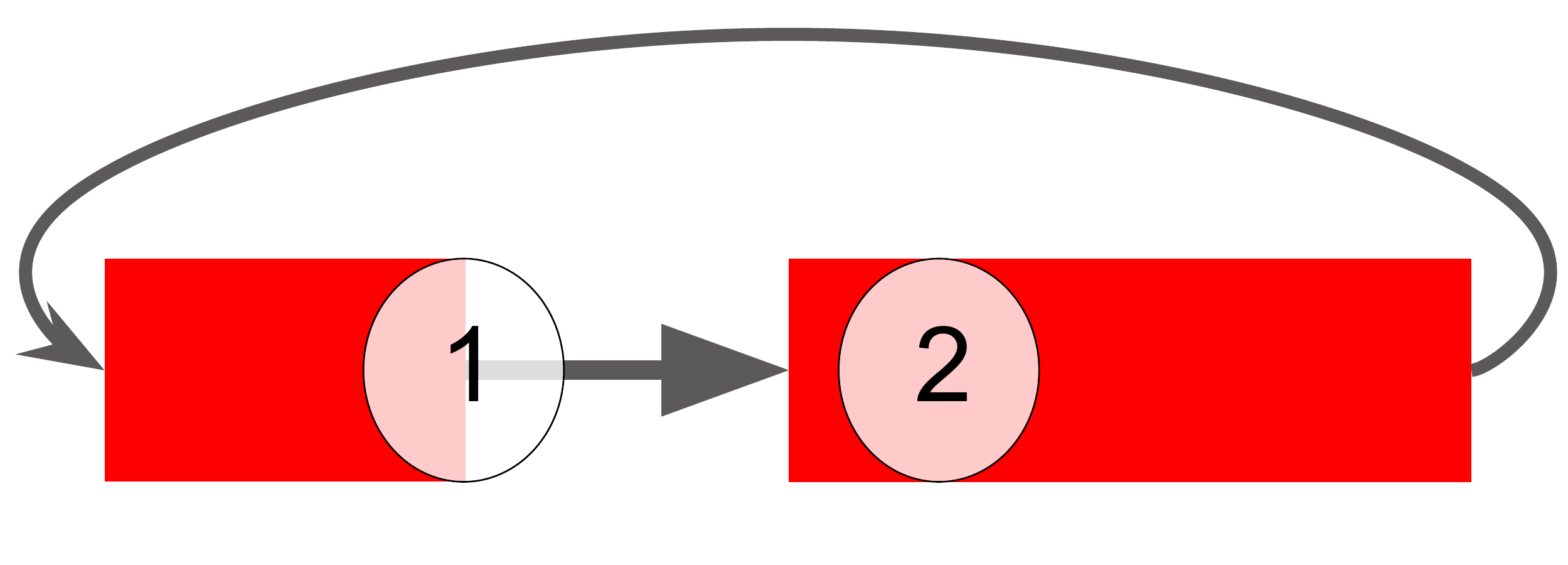} 
		   \label{placart3}
   \end{subfigure}
   \caption{(Color Online) A schematic to illustrate the criteria used for constructing the phase diagram from  PLA:  (a) boundary between amplitude death and  out of phase oscillations (b) condition for in-phase oscillations, and (c) condition for {\it all} oscillations being in phase.}
   
   \end{figure}

\begin{figure}[htp] %  figure placement: here, top, bottom, or page
   \includegraphics[width=.95\linewidth]{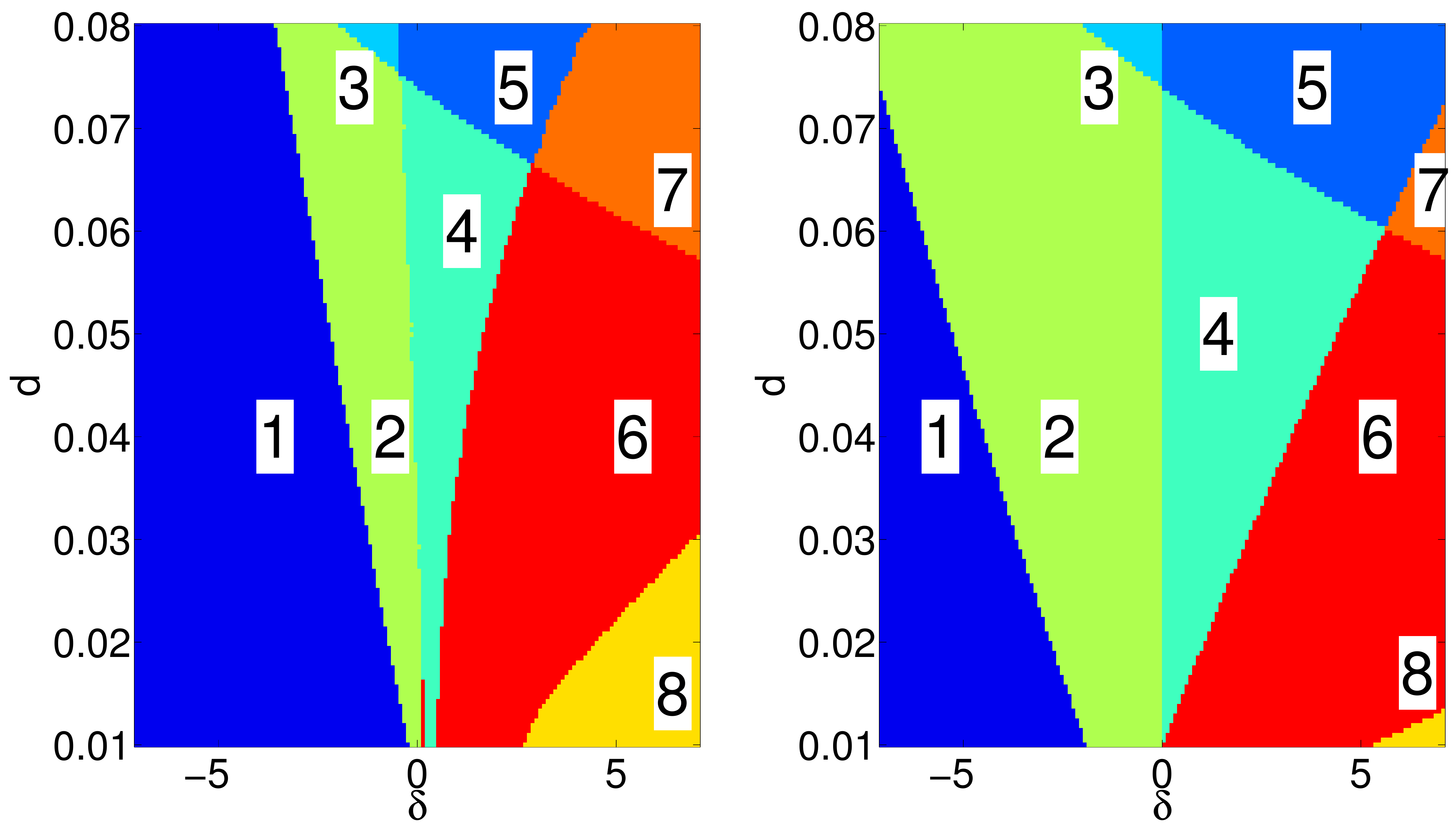} 
   \caption{(Color Online) Left - Phase diagram constructed from numerical simulations. Right - Phase diagram constructed from PLA.  Significant differences include the sizes of 2 and  8.   Regions 2-7 are $d-\delta$ intervals  where we expect  out of phase oscillations. {The left bound on this region obtained from PLA is a lower bound to  the numerical results, since we assumed that one of the Brusselators would start at $\alpha$ instead of $\alpha-d$.  The PLA, therefore, underestimates the distance to be travelled along the nullcline, leading to an underestimation of the coupling strength $d$. Oscillator death occurs in  regions 3, 5, and 7.  The PLA estimation of these regions should be exact.   Regions 6-8 are where in-phase oscillations can occur.   The full nonlinear equations increase the stability of in-phase oscillations since for a finite $c$ the jump form $\alpha$ to $2\alpha$ is not  instantaneous.}}
   \label{fig:phasecomp}
\end{figure}

Since PLA provides a very good approximation to the dynamics at large $c$, we use it to construct a phase diagram in $\delta-d$ space by defining boundaries between the states listed in Table \ref{tab:phases}, and comparing it to the one obtained from numerical results. The finally comparison can be seen in Fig. \ref{fig:phasecomp}
We first investigate the boundary between amplitude death and out of phase oscillations. We note  from linear stability analysis that for $\delta<0$ the fixed point is stable and therefore amplitude death can occur. Next we examine the bound on when out of phase oscillations can occur. To do this we imagine both Brusselators to be on the $L$ branch: one at ($u_1=\beta$) and one just about to jump from $u_2=\alpha$ to $u_2=2\alpha$.
 
\begin{align}
\dot{u}_1=&-\frac{1}{\eta}[1+\beta-u_{max}+d(\alpha-\beta)]\\
\dot{u}_2=&-\frac{1}{\eta}[1-\alpha-d(\alpha-\beta)]\label{eq:ampoop}
\end{align}

If $\dot{u}_2<0$ then the Brusselators will always fall to the fixed point. If on the other hand $\dot{u}_2>0$ then there exist stable out of phase oscillations.
The phase boundary between between amplitude death, and multistability of amplitude death and out of phase oscillations is, therefore,  determined by  the condition $\dot{u}_2=0$.
It is then apparent from Eq. \ref{eq:ampoop} that for $d>\frac{1-\alpha}{\alpha-\beta}$ the coupling can cause the pair of Brusselators to oscillate.
We now examine the conditions for the occurrence of in-phase oscillations.  We assume the oscillators start out in phase and we examine  the behavior around the $u=\alpha$ to $u=2\alpha$ jump. We assume the oscillators progress along the nullcline with $u_1 \simeq u_2$ but one oscillator is infinitesimally ahead of the other and jumps from $\alpha$ to $2\alpha$ first. Without loss of generality we say that $u_2$ jumps before $u_1$. Then we have the oscillators at the coordinates $(u_1,u_2)=(\alpha,2\alpha)$. At these coordinates the derivatives are
\begin{align}
\dot{u}_1=&-\frac{1}{\eta}(1-\alpha+d\alpha)\\
\dot{u}_2=&1-d\alpha
\end{align}
Now we can see that if  $d<1-\frac{1}{\alpha}$,  $\dot{u}_1$ is positive implying that the coupling is not strong enough to keep $u_1$ in $L$, and therefore,  the Brusselators can oscillate in phase. 

%When we move to higher numbers of Brusselators the perception is the same. We find assume the Brusselators are in phase and find that if $d>1-\frac{1}{\alpha}$ a Brusselator inhibits its nearest neighbors and can't oscillate in phase. 
To find a good bound on where all oscillations must be in phase we start with the Brusselators out of phase and see when the coupling leads to in phase behavior.  We first look at the single Brusselator dynamics. The assumption here is that time scales are not affected stongly by the coupling in the regime where the coupling is small enough that oscillations are in phase.  We calculate the time it  takes to go along the branch $L$. This is done by solving the first order differential equation with initial conditions at $u(0)=\beta$ and we find that the Brusselator is ready to jump at $u(t_1)=\alpha$ where to first order in $\frac{1}{c}$
\begin{equation}
t_1=\log{\frac{c^2}{2(1+\delta)}}
\end{equation}
The velocity on the right branch is $\dot{u}=1$ so if the 2 Brusselators start $(u_1=\beta, u_2=\alpha)$ they will progress to approximately $\left(u_1\approx\alpha,u_2\approx2\alpha+\log(\frac{c^2}{2(1+\delta)})\right)$. With derivatives

\begin{widetext}
\begin{align}
\dot{u}_1=&-\frac{1}{\eta}\left(1-\alpha +d\left[\alpha+\log(\frac{c^2}{2(1+\delta)})\right]\right)\\
\dot{u}_2=&1-d\left[\alpha+\log(\frac{c^2}{2(1+\delta)})\right]\nonumber
\end{align}
\end{widetext}

Now we check when the coupling is strong enough to keep Brusselator 1 from jumping at these coordinates. We find that for $d<\frac{\alpha-1}{\alpha+\log(\frac{c^2}{2(1+\delta)})}$ the coupling is not strong enough to keep both Brusselators off the right branch, even if they start out of phase. When they are on the right branch the attractive coupling can then pull Brusselators together to form in phase oscillations. 
 
When we combine the above criteria with condition for oscillator death discussed in section \ref{sec:oscdeath}, a phase portrait emerges, which is remarkably  similar to the one obtained from numerical simulations, as seen from Fig. \ref{fig:phasecomp}

%Notably when coupling constant is strong enough, namely $d>1-\frac{1}{\alpha}$ the jump from $L$ to $R$ is enough to cause an oscillator in $L$ to stay in $L$ and not be able to jump to $R$ until both oscillators are back in $L$.
%\begin{figure}[H] %  figure placement: here, top, bottom, or page
%	\centering
%	\begin{subfigure}[t]{0.45\linewidth}
%		   \includegraphics[width=\linewidth]{ipPLACart} 
%   \caption{A cartoon depicting a $u_{1}$, $u_{2}$ phase portrait depicting an in phase trajectory. This is just the line $u_1=u_2$. The black line mark the discontinuities in u space between $\alpha$ and $2\alpha$.}
%   \end{subfigure}
%   \quad
%        \begin{subfigure}[t]{0.45\linewidth}
%   \includegraphics[width=\linewidth]{oofPLACart} 
%   \caption{ $u_{1}$, $u_{2}$ phase portrait depicting an out of phase trajectory. The system spends most of its time in the parts of phase space where one oscillator is on $L$ very close to the jump while the other continues on $R$. }
%   \end{subfigure}
%   \caption{A cartoon depicting a Phase portraits of in phase and out of phase modes. The black line mark the discontinuities in u space between $\alpha$ and $2\alpha$.}
%\end{figure}
By greatly simplifying the dynamics we have gained some insight into the behavior of the full system when the nonlinearity is strong.

\section{Conclusion}{
We have presented a detailed analysis of pattern formation in coupled Brusselators in the regime of high nonlinearity.  This regime is characterized by fast inhibitor dynamics.   Coupling via this fast species leads to preference for out-of-phase  oscillations: a feature that is absent in the regime of low or moderate nonlinearity.  Numerical studies of rings of $N_r$ coupled oscillators lead to a rich phase diagram with  regions of oscillator death, in and out of phase oscillations, and multistability. Analysis of two coupled Brusselators using a piecewise linear approximation provides a detailed picture of the dynamics that leads to out-of-phase oscillations.  The crucial observation is that in the limit of very strong nonlinearity the limit cycle can be broken down into two branches in one of which the coupling is repulsive and  in the other the coupling is attractive.   A phase diagram based on the PLA   is found to be in good agreement with numerical results.   
%An study of the Brusselator has been carried out, with emphasis on large coefficient of the nonlinear term. First a change of variables was performed to emphasize the time scale separation. A canard explosion was then numerically show to be present. Then a standard Turing analysis was carried out for a general ring of N Brusselators coupled though the inhibitor. The Turing analysis suggested that the Brusselators will oscillate in phase.

%This rich attractor space is to a large extent a consequence of the geometrical frustration  that arises because of the preference for oscillators to synchronize out of phase.

%This prediction of the Turing analysis is was then contrasted with the results numerically obtained from rings of $N=2..5$ Brusselators. These rings exhibited a variety of modes of oscillation including out of phase oscillations and modes that relived the frustrated geometry of the ring by forming an in phase pair. An exploration of phase space of coupling and bifurcation parameter was then carried out for $N=2$ and $N=5$ rings of Brusselators.
%
%Finally a piecewise linear approximation was used to explain some of the phenomena observed numerically. In the limit of very strong nonlinearity is was show that the limit cycle can be broken down into two branches, with where the coupling is repulsive and one where the  coupling is attractive. Using the approximation an choosing specific initial conditions an approximate phase diagram was constructed akin to the one found though numerical means.

Numerical analysis  for a ring of $5$ oscillators yields patterns in which the {\it frequencies} of the oscillators synchronize with integer frequency ratios.   For rings with fewer than $5$ oscillators, the diversity of patterns is characterized by their phase relationships since the frequencies of all oscillators are identical.   The PLA analysis suggests that one can reduce the coupled dynamics of strongly nonlinear Brusselators to a discrete-time map, which are know to exhibit frequency locking\cite{Bak_Devils}.   This analysis and a detailed numerical study of the  complete phase diagram for the $N_r=3$, and $4$ rings will be the subject of future research. 

\vspace{0.1in}
\section{Acknowledgements}  We acknowledge extensive discussions with Irving Epstein, Nate Tompkins and Seth Fraden, and Royce Zia.  This was work was partially supported by the Brandeis MRSC (NSF-DMR0820492) and the Brandeis IGERT (NSF-DGE1068620). MG's work was supported in part by  NSF-DMR-1244666.  BC's work was supported in part by NSF- PHYS-1066293 and the hospitality of the Aspen Center for Physics.}
%\end{multicols}
\bibliographystyle{aipnum4-1}
\bibliography{Brusselator_Paper}
%\subsection{}

\end{document}